\documentclass[journal,draftclsnofoot,onecolumn]{IEEEtran}
\usepackage{graphicx}
\usepackage{amssymb}

\usepackage{caption}
\renewcommand{\theequation}{\arabic{section}.\arabic{equation}}
\newtheorem{theorem}{Theorem}

\newtheorem{lemma}{Lemma}
\newtheorem{definition}{Definition}
\newtheorem{corollary}{Corollary}
\newtheorem{remark}{Remark}

\begin{document}

\title{Information-Theoretical Security for Several Models of Multiple-Access Channel}

\author{Bin~Dai,
        A.~J.~Han~Vinck,~\IEEEmembership{Fellow,~IEEE,}
        Zhuojun~Zhuang,
        and Yuan~Luo
\thanks{B. Dai is with the
Computer Science and Engineering Department,
Shanghai Jiao Tong University, and the
Institute for Experimental Mathematics, Duisburg-Essen University,
Ellernstr.29, 45326 Essen, Germany e-mail: daibinsjtu@gmail.com.}
\thanks{A. J. Han Vinck is with the
Institute for Experimental Mathematics, Duisburg-Essen University,
Ellernstr.29, 45326 Essen, Germany
e-mail: vinck@iem.uni-due.de.}
\thanks{Z. Zhuang is with the
Computer Science and Engineering Department,
Shanghai Jiao Tong University,
Shanghai 200240, China e-mail: zhuojunzzj@sjtu.edu.cn.}
\thanks{Y. Luo is with the
Computer Science and Engineering Department,
Shanghai Jiao Tong University,
Shanghai 200240, China e-mail: luoyuan@cs.sjtu.edu.cn.}
}

\maketitle

\begin{abstract}
Several security models of multiple-access channel (MAC) are investigated. First, we study the degraded MAC
with confidential messages, where two users transmit their confidential messages (no common message) to a destination,
and each user obtains a degraded version of the output of the MAC. Each user views the other user as a eavesdropper, and wishes
to keep its confidential message as secret as possible from the other user. Measuring each user's uncertainty about the other user's
confidential message by equivocation, the inner and outer bounds on the capacity-equivocation region for this model have been provided.
The result is further
explained via the binary and Gaussian examples.

Second, the discrete memoryless multiple-access wiretap channel (MAC-WT) is studied, where two users transmit their corresponding
confidential messages (no common message) to a legitimate receiver, while an additional wiretapper wishes to obtain the messages via a wiretap channel.
This new model is considered into two cases: the general MAC-WT with cooperative encoders,
and the degraded MAC-WT with non-cooperative encoders. The capacity-equivocation region is totally determined for
the cooperative case, and inner and outer bounds on the capacity-equivocation region are provided for
the non-cooperative case. For both cases, the results are further explained via the binary examples.

\end{abstract}


\section{Introduction \label{secI}}
Transmission of confidential messages has been studied in the literature of several classes of channels.
Wyner, in his well-known paper on the wiretap channel \cite{Wy}, studied the problem that how to transmit
the confidential messages to the legitimate receiver via a degraded broadcast channel, while keeping the wiretapper
as ignorant of the messages as possible. Measuring the uncertainty of the wiretapper by equivocation, the capacity-equivocation
region was established. Furthermore, the secrecy capacity was also established, which provided the maximum transmission rate with perfect secrecy.
After the publication of Wyner's work, Csisz$\acute{a}$r and K\"{o}rner \cite{CK} investigated a more
general situation: the broadcast channels with confidential messages (BCC). In this model,
a common message and a confidential message were sent through a general broadcast channel. The common message was assumed to be decoded correctly
by the legitimate receiver and the wiretapper, while the confidential message was only allowed to be obtained by the legitimate receiver.
This model is also a generalization of \cite{KM}, where no confidentiality condition is imposed. The capacity-equivocation region
and the secrecy capacity region of BCC \cite{CK}
were totally determined, and the results were also a generalization of those in \cite{Wy}.
Based on Wyner's work, Leung-Yan-Cheong and Hellman studied the Gaussian
wiretap channel(GWC) \cite{CH}, and showed that its secrecy
capacity was the difference between the main channel capacity and the overall wiretap channel
capacity (the cascade of main channel and wiretap channel). Some other related works on the wiretap channel (including feedback, side
information and secret key) can be found in
\cite{AC}, \cite{AFJK}, \cite{LGP}, \cite{Me}, \cite{Ch}, \cite{MVL}.

\begin{figure}[htb]
\centering
\includegraphics[scale=0.6]{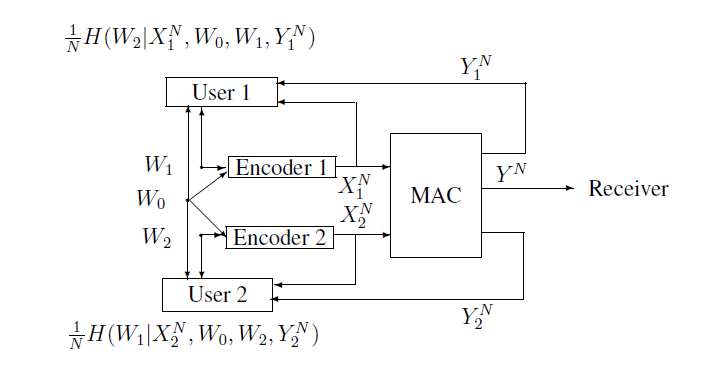}
\caption{MAC with confidential messages}
\label{f1}
\end{figure}

Recently, the information-theoretical security for other multi-user communication systems has been investigated.
The relay channel with confidential messages was studied in \cite{LG2}, and the interference channel with confidential messages was studied in \cite{LMSY}.
For the multiple-access channel, the security problems are split into two directions. The first is that
two users wish to transmit their corresponding messages to a destination, and meanwhile, they also receive
the channel output. Each user treats the other user as a wiretapper, and wishes to keep its confidential message as secret as possible
from the wiretapper. This model is usually called the MAC with confidential messages, and it was studied by \cite{LP}, see Figure \ref{f1}.
An inner bound on the capacity-equivocation region is provided for the model of Figure \ref{f1}, and the
capacity-equivocation region is still not known. Furthermore, for the model of MAC with one confidential message \cite{LP}, both inner and outer bounds
on capacity-equivocation region are derived. Moreover, for the degraded MAC with one confidential message, the
capacity-equivocation region is totally determined.

\begin{figure}[htb]
\centering
\includegraphics[scale=0.6]{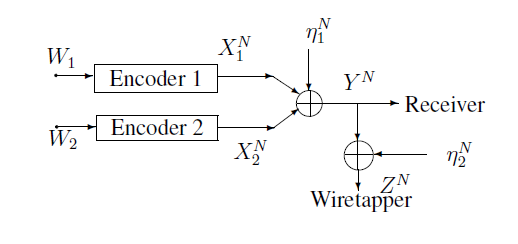}
\caption{Gaussian multiple-access wiretap channel}
\label{f2}
\end{figure}

The second is that an additional wiretapper has access to the MAC output via a wiretap channel, and therefore, how to keep the confidential
messages of the two users as secret as possible from the additional wiretapper is the main concern of the system designer.
This model is usually called the multiple-access wiretap channel (MAC-WT). The Gaussian MAC-WT was investigated by \cite{TY2},
see Figure \ref{f2}. An inner bound on the capacity-equivocation region is provided for the Gaussian MAC-WT. Other related works
on MAC-WT can be found in \cite{TY1}, \cite{EU}.

\begin{figure}[htb]
\centering
\includegraphics[scale=0.6]{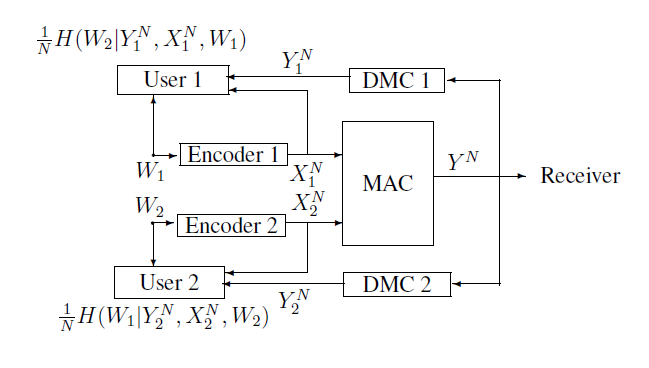}
\caption{Degraded MAC with confidential messages}
\label{f3}
\end{figure}

In this paper, firstly we study a special case of Figure \ref{f1}, where two users wish to transmit their confidential messages (no common message)
to a destination, and meanwhile, they also receive a degraded version of the channel output, see Figure \ref{f3}.
Each user wishes to keep its confidential message as secret as possible
from the other user. Measuring each user's uncertainty about the other one's confidential message by equivocation, the inner and outer bounds on the
capacity-equivocation
region are provided for this model. Then, as examples, we establish the inner and outer bounds on the capacity-equivocation regions for the Gaussian and binary cases
 of Figure \ref{f3}.

\begin{figure}[htb]
\centering
\includegraphics[scale=0.6]{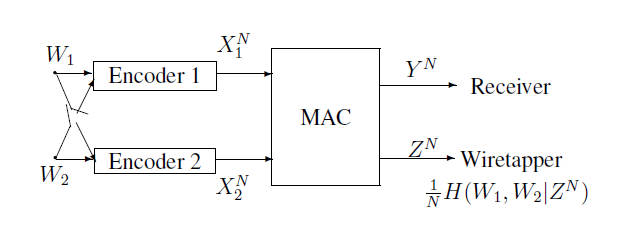}
\caption{Multiple-access wiretap channel with cooperative (or non-cooperative) encoders}
\label{f4}
\end{figure}

Secondly we study the discrete memoryless multiple-access wiretap channel (MAC-WT), see Figure \ref{f4}.
The model of Figure \ref{f4} is considered into two cases: MAC-WT with cooperative encoders, and degraded MAC-WT with non-cooperative encoders.
For the MAC-WT with cooperative encoders, the capacity-equivocation
region is determined. Furthermore, if the received symbols for the wiretapper is a degraded version of the symbols
for the legitimate receiver (usually called degraded MAC-WT with cooperative encoders),
we also establish the capacity-equivocation region for this special case.
For the degraded MAC-WT with non-cooperative encoders, inner and outer bounds on the capacity-equivocation region
are provided.
Finally, as examples, we give the capacity-equivocation region for the binary degraded MAC-WT with cooperative encoders, and the secrecy capacity region
for the binary degraded MAC-WT with non-cooperative encoders.

In this paper, random variab1es, sample values and
alphabets are denoted by capital letters, lower case letters and calligraphic letters, respectively.
A similar convention is applied to the random vectors and their sample values.
\textbf{For example, $U^{N}$ denotes a random $N$-vector $(U_{1},...,U_{N})$,
and $u^{N}=(u_{1},...,u_{N})$ is a specific vector value in $\mathcal{U}^{N}$
that is the $N$th Cartesian power of $\mathcal{U}$.
$U_{i}^{N}$ denotes a random $N-i+1$-vector $(U_{i},...,U_{N})$,
and $u_{i}^{N}=(u_{i},...,u_{N})$ is a specific vector value in $\mathcal{U}_{i}^{N}$.}
Let $p_{V}(v)$ denote the probability mass function $Pr\{V=v\}$.
Throughout the paper, the logarithmic function is to the base 2.

The organization of this paper is as follows. In Section II, the capacity-equivocation region
and the secrecy capacity region of the model of Figure \ref{f3} are determined in Theorem \ref{T1} and Remark \ref{R1}, respectively.
Then, as two examples, the capacity-equivocation region and the secrecy capacity region for the Gaussian and binary cases of Figure \ref{f3}
are shown in Section III.
In Section IV, the capacity-equivocation region and the secrecy capacity region for
the model of Figure \ref{f4} with cooperative encoders are determined in Theorem \ref{T4} and Remark \ref{R4}, respectively.
The results of the degraded case for the model of Figure \ref{f4} with cooperative encoders are also shown in Section IV.
The inner and outer bounds on the capacity-equivocation region for the
model of Figure \ref{f4} with non-cooperative encoders are shown in Section V.
In Section VI, we will show the binary examples about the model of Figure \ref{f4} with cooperative or non-cooperative encoders.
Final conclusions are provided in Section VII.

\section{Degraded MAC with Confidential Messages\label{secII}}

In this section, a description of the  model of Figure \ref{f3} is given by Definition \ref{def1} to Definition \ref{def3}.
The capacity-equivocation region $\mathcal{R^{(A)}}$ composed of all achievable $(R_{1}, R_{2}, R_{e1}, R_{e2})$ tuples in the model of Figure \ref{f3} is
characterized in Theorem \ref{T1}, where the achievable $(R_{1}, R_{2}, R_{e1}, R_{e2})$ tuple is defined in Definition \ref{def4}.

\begin{definition}(\textbf{Encoders})\label{def1}
The confidential messages $W_{1}$ and $W_{2}$ take values
in $\mathcal{W}_{1}$ and $\mathcal{W}_{2}$, respectively. $W_{1}$ and $W_{2}$ are independent and uniformly
distributed over their ranges. Since each encoder is a wiretapper for the other
encoder, the cooperation between the encoders is not allowed.
The input and output of encoder 1 are $W_{1}$ and $X_{1}^{N}$, respectively.
Similarly, the input and output of encoder 2 are $W_{2}$ and $X_{2}^{N}$, respectively.
We assume that the encoders are stochastic encoders, i.e., the encoder
$f_{i}^{N}$ ($i=1,2$) is a matrix of conditional probabilities $f_{i}^{N}(x_{i}^{N}|w_{i})$,
where $x_{i}^{N}\in \mathcal{X}_{i}^{N}$, $w_{i}\in \mathcal{W}_{i}$, and $f_{i}^{N}(x_{i}^{N}|w_{i})$ is the probability that the message
$w_{i}$ is encoded as the channel input $x_{i}^{N}$.
Note that $W_{1}$ and $X_{2}^{N}$ are independent, and $W_{2}$ is independent of  $X_{1}^{N}$.

The transmission rates of the confidential messages are $\frac{\log\parallel \mathcal{W}_{1}\parallel}{N}$ and
$\frac{\log\parallel \mathcal{W}_{2}\parallel}{N}$.
\end{definition}

\begin{definition}(\textbf{Channels})\label{def2}
The MAC is a DMC with finite input alphabet
$\mathcal{X}_{1}\times \mathcal{X}_{2}$, finite output alphabet $\mathcal{Y}$, and
transition probability $Q_{1}(y|x_{1},x_{2})$, where $x_{1}\in \mathcal{X}_{1},x_{2}\in \mathcal{X}_{2},y\in
\mathcal{Y}$. $Q_{1}(y^{N}|x_{1}^{N},x_{2}^{N})=\prod_{n=1}^{N}Q_{1}(y_{n}|x_{1,n},x_{2,n})$.
The inputs of the MAC are $X_{1}^{N}$ and $X_{2}^{N}$, while the output is $Y^{N}$.

User $i$ ($i=1,2$) has access to the output of the MAC via channel $i$. Channel $i$ is a DMC with input $Y^{N}$ and
output $Y_{i}^{N}$. User 2's equivocation about $W_{1}$ is defined as
\begin{equation}\label{e201}
\Delta_{1}=\frac{1}{N}H(W_{1}|Y_{2}^{N},W_{2},X_{2}^{N})\stackrel{(a)}=\frac{1}{N}H(W_{1}|Y_{2}^{N},X_{2}^{N}),
\end{equation}
and user 1's equivocation about $W_{2}$ is defined as
\begin{equation}\label{e202}
\Delta_{2}=\frac{1}{N}H(W_{2}|Y_{1}^{N},W_{1},X_{1}^{N})\stackrel{(b)}=\frac{1}{N}H(W_{2}|Y_{1}^{N},X_{1}^{N}),
\end{equation}
where (a) and (b) are from $W_{2}\rightarrow (Y_{2}^{N},X_{2}^{N})\rightarrow W_{1}$ and
$W_{1}\rightarrow (Y_{1}^{N},X_{1}^{N})\rightarrow W_{2}$.

\end{definition}

\begin{definition}(\textbf{Decoder})\label{def3}
The decoder is a mapping $f_{D}: \mathcal{Y}^{N}\rightarrow \mathcal{W}_{1}\times \mathcal{W}_{2}$,
with input $Y^{N}$ and outputs $\widehat{W}_{1}$, $\widehat{W}_{2}$. Let $P_{e}$ be the error probability of the receiver
, and it is
defined as $Pr\{(W_{1},W_{2})\neq (\widehat{W}_{1}, \widehat{W}_{2})\}$.
\end{definition}

\begin{definition}(\textbf{Achievable $(R_{1}, R_{2}, R_{e1},R_{e2})$ tuple in the model of Figure \ref{f3}})\label{def4}
A tuple $(R_{1}, R_{2}, R_{e1},R_{e2})$ (where $R_{1}, R_{2}, R_{e1}, R_{e2}>0$) is called
achievable if, for any $\epsilon>0$ (where $\epsilon$ is an arbitrary small positive real number
and $\epsilon\rightarrow 0$), there exists a channel
encoder-decoder $(N, \Delta_{1}, \Delta_{2}, P_{e})$ such that
\begin{equation}\label{e203}
\lim_{N\rightarrow \infty}\frac{\log\parallel \mathcal{W}_{1}\parallel}{N}= R_{1} \label{e203a},
\lim_{N\rightarrow \infty}\frac{\log\parallel \mathcal{W}_{2}\parallel}{N}= R_{2} \label{e203b},
\lim_{N\rightarrow \infty}\Delta_{1}\geq R_{e1} \label{e203c},
\lim_{N\rightarrow \infty}\Delta_{2}\geq R_{e2} \label{e203d}, P_{e}\leq \epsilon\label{e203e}.
\end{equation}
\end{definition}

\textbf{The capacity-equivocation region $\mathcal{R}^{(A)}$ is a set composed of all achievable $(R_{1}, R_{2}, R_{e1},R_{e2})$ tuples.
The inner and outer bounds on the capacity-equivocation region $\mathcal{R}^{(A)}$ are provided in Theorem \ref{T1} and Theorem \ref{T1x}, respectively,
and they are proved in Appendix \ref{appen1b} and Appendix \ref{appen1}.}

\begin{theorem}\label{T1}
\textbf{(Inner bound)} A single-letter characterization of the region $\mathcal{R}^{(Ai)}$ is as follows,
\begin{eqnarray*}
&&\mathcal{R}^{(Ai)}=\{(R_{1}, R_{2}, R_{e1},R_{e2}): 0\leq R_{e1}\leq R_{1}, 0\leq R_{e2}\leq R_{2},\\
&&0\leq R_{1}\leq I(X_{1};Y|X_{2}),\\
&&0\leq R_{2}\leq I(X_{2};Y|X_{1}), \\
&&R_{1}+R_{2}\leq I(X_{1},X_{2};Y),\\
&&R_{e1}\leq I(X_{1};Y|X_{2})-I(X_{1};Y_{2}|X_{2}),\\
&&R_{e2}\leq I(X_{2};Y|X_{1})-I(X_{2};Y_{1}|X_{1}),\\
&&R_{e1}+R_{e2}\leq I(X_{1},X_{2};Y)-I(X_{1};Y_{2}|X_{2})-I(X_{2};Y_{1}|X_{1}),\\
&&R_{e1}+R_{2}\leq I(X_{1},X_{2};Y)-I(X_{1};Y_{2}|X_{2}),\\
&&R_{e2}+R_{1}\leq I(X_{1},X_{2};Y)-I(X_{2};Y_{1}|X_{1})\},
\end{eqnarray*}
where
$(X_{1},X_{2})\rightarrow Y\rightarrow (Y_{1},Y_{2})$ and $\mathcal{R}^{(Ai)}\subseteq \mathcal{R}^{(A)}$.
\end{theorem}

\begin{remark}\label{R1}
There are some notes on Theorem \ref{T1}, see the following.
\begin{itemize}

\item The region $\mathcal{R^{(A)}}$ is convex, and the proof is directly obtained by introducing a time sharing random variable into Theorem \ref{T1},
and therefore, we omit the proof here.

\item Note that Theorem \ref{T1} indicates a tradeoff between the two equivocations $R_{e1}$ and $R_{e2}$, i.e.,
$R_{e1}+R_{e2}\leq I(X_{1},X_{2};Y)-I(X_{1};Y_{2}|X_{2})-I(X_{2};Y_{1}|X_{1})$.

\item The achievable secrecy region $\mathcal{C}_{s}^{(Ai)}$ is the set of pairs $(R_{1},R_{2})$ such that
$(R_{1}, R_{2}, R_{e1}=R_{1},R_{e2}=R_{2})\in \mathcal{R}^{(Ai)}$.
\begin{corollary}
\begin{eqnarray*}
&&\mathcal{C}_{s}^{(Ai)}=\{(R_{1}, R_{2}): R_{1}\leq I(X_{1};Y|X_{2})-I(X_{1};Y_{2}|X_{2}), \\
&&R_{2}\leq I(X_{2};Y|X_{1})-I(X_{2};Y_{1}|X_{1}),\\
&&R_{1}+R_{2}\leq I(X_{1},X_{2};Y)-I(X_{1};Y_{2}|X_{2})-I(X_{2};Y_{1}|X_{1})\}.
\end{eqnarray*}
\end{corollary}
\begin{IEEEproof}
Corollary 1 is easy to be checked by substituting $R_{e1}=R_{1}$ and $R_{e2}=R_{2}$ into $\mathcal{R}^{(Ai)}$.
\end{IEEEproof}

\end{itemize}
\end{remark}

\begin{theorem}\label{T1x}
\textbf{(Outer bound)} A single-letter characterization of the region $\mathcal{R}^{(Ao)}$ is as follows,
\begin{eqnarray*}
&&\mathcal{R}^{(Ao)}=\{(R_{1}, R_{2}, R_{e1},R_{e2}): 0\leq R_{e1}\leq R_{1}, 0\leq R_{e2}\leq R_{2},\\
&&0\leq R_{1}\leq I(X_{1};Y|X_{2}),\\
&&0\leq R_{2}\leq I(X_{2};Y|X_{1}), \\
&&R_{1}+R_{2}\leq I(X_{1},X_{2};Y),\\
&&R_{e1}\leq I(X_{1};Y|X_{2})-I(X_{1};Y_{2}|X_{2}),\\
&&R_{e2}\leq I(X_{2};Y|X_{1})-I(X_{2};Y_{1}|X_{1})\},
\end{eqnarray*}
where
$(X_{1},X_{2})\rightarrow Y\rightarrow (Y_{1},Y_{2})$ and $\mathcal{R}^{(A)}\subseteq \mathcal{R}^{(Ao)}$.
\end{theorem}

\begin{remark}\label{R1x}
There are some notes on Theorem \ref{T1x}, see the following.
\begin{itemize}

\item The region $\mathcal{R^{(A)}}$ is convex, and the proof is omitted here.

\item The outer bound $\mathcal{C}_{s}^{(Ao)}$ on the secrecy capacity region is the set of pairs $(R_{1},R_{2})$ such that
$(R_{1}, R_{2}, R_{e1}=R_{1},R_{e2}=R_{2})\in \mathcal{R}^{(Ao)}$.
\begin{corollary}
\begin{eqnarray*}
&&\mathcal{C}_{s}^{(Ao)}=\{(R_{1}, R_{2}): R_{1}\leq I(X_{1};Y|X_{2})-I(X_{1};Y_{2}|X_{2}), \\
&&R_{2}\leq I(X_{2};Y|X_{1})-I(X_{2};Y_{1}|X_{1}),\\
&&R_{1}+R_{2}\leq I(X_{1},X_{2};Y)\}.
\end{eqnarray*}
\end{corollary}
\begin{IEEEproof}
Corollary 2 is easy to be checked by substituting $R_{e1}=R_{1}$ and $R_{e2}=R_{2}$ into $\mathcal{R}^{(Ao)}$.
\end{IEEEproof}

\end{itemize}
\end{remark}

\section{Gaussian and Binary MACs with Confidential Messages\label{sec2.2}}

\subsection{The Gaussian Case of the Model of Figure \ref{f3}\label{sub31}}

In this subsection, we study the Gaussian case of Figure \ref{f3}, where the channel input-output relationships
at each time instant $i$ ($1\leq i\leq N$) are given by
\begin{equation}\label{e301}
Y_{i}=X_{1,i}+X_{2,i}+Z_{i},
\end{equation}
\begin{equation}\label{e302}
Y_{1,i}=X_{1,i}+X_{2,i}+Z_{i}+Z_{1,i},
\end{equation}
and
\begin{equation}\label{e303}
Y_{2,i}=X_{1,i}+X_{2,i}+Z_{i}+Z_{2,i},
\end{equation}
where $Z_{i}\sim \mathcal{N}(0,N_{0})$, $Z_{1,i}\sim \mathcal{N}(0,N_{1})$ and $Z_{2,i}\sim \mathcal{N}(0,N_{2})$.
The random vectors $Z^{N}$, $Z_{1}^{N}$ and $Z_{2}^{N}$ are independent with i.i.d. components.
The channel inputs $X_{1}^{N}$ and $X_{2}^{N}$ are subject to the average power constraints $P_{1}$ and $P_{2}$, respectively,
i.e.,
\begin{equation}\label{e304}
\frac{1}{N}\sum_{i=1}^{N}E[X^{2}_{1,i}]\leq P_{1},\quad\quad\quad \frac{1}{N}\sum_{i=1}^{N}E[X^{2}_{2,i}]\leq P_{2}.
\end{equation}

The following Theorem \ref{T2} and Theorem \ref{T2x} provide inner and outer bounds on the capacity-equivocation region of Gaussian
MAC with confidential messages.

\begin{theorem}\label{T2}
For the Gaussian case of Figure \ref{f3}, the inner bound $\mathcal{R}^{(Bi)}$ on the capacity-equivocation region $\mathcal{R^{(B)}}$ is given by
\begin{equation}
\mathcal{R}^{(Bi)}=\bigcup_{0\leq \alpha\leq 1, 0\leq \beta\leq 1}
\left\{
\begin{array}{ll}
(R_{1}, R_{2}, R_{e1},R_{e2}):\\
0\leq R_{1}\leq \frac{1}{2}\log(1+\frac{\alpha P_{1}}{N_{0}})\\
0\leq R_{2}\leq \frac{1}{2}\log(1+\frac{\beta P_{2}}{N_{0}})\\
R_{1}+R_{2}\leq \frac{1}{2}\log(1+\frac{P_{1}+P_{2}+2\sqrt{(1-\alpha)P_{1}P_{2}}}{N_{0}})\\
R_{1}+R_{2}\leq \frac{1}{2}\log(1+\frac{P_{1}+P_{2}+2\sqrt{(1-\beta)P_{1}P_{2}}}{N_{0}})\\
0\leq R_{e1}\leq R_{1}\\
0\leq R_{e2}\leq R_{2}\\
R_{e1}\leq \frac{1}{2}\log(1+\frac{\alpha P_{1}}{N_{0}})-\frac{1}{2}\log(1+\frac{\alpha P_{1}}{N_{0}+N_{2}})\\
R_{e2}\leq \frac{1}{2}\log(1+\frac{\beta P_{2}}{N_{0}})-\frac{1}{2}\log(1+\frac{\beta P_{2}}{N_{0}+N_{1}})\\
R_{e1}+R_{e2}\leq \frac{1}{2}\log(1+\frac{P_{1}+P_{2}+2\sqrt{(1-\alpha)P_{1}P_{2}}}{N_{0}})-\frac{1}{2}\log(1+\frac{\alpha P_{1}}{N_{0}+N_{2}})-\frac{1}{2}\log(1+\frac{\beta P_{2}}{N_{0}+N_{1}})\\
R_{e1}+R_{e2}\leq \frac{1}{2}\log(1+\frac{P_{1}+P_{2}+2\sqrt{(1-\beta)P_{1}P_{2}}}{N_{0}})-\frac{1}{2}\log(1+\frac{\alpha P_{1}}{N_{0}+N_{2}})-\frac{1}{2}\log(1+\frac{\beta P_{2}}{N_{0}+N_{1}})\\
R_{e1}+R_{2}\leq \frac{1}{2}\log(1+\frac{P_{1}+P_{2}+2\sqrt{(1-\alpha)P_{1}P_{2}}}{N_{0}})-\frac{1}{2}\log(1+\frac{\alpha P_{1}}{N_{0}+N_{2}})\\
R_{e2}+R_{1}\leq \frac{1}{2}\log(1+\frac{P_{1}+P_{2}+2\sqrt{(1-\beta)P_{1}P_{2}}}{N_{0}})-\frac{1}{2}\log(1+\frac{\beta P_{2}}{N_{0}+N_{1}}).
\end{array}
\right\}.
\end{equation}
\end{theorem}
\begin{IEEEproof}
See Appendix \ref{appen3}.
\end{IEEEproof}

\begin{corollary} The inner bound on the secrecy capacity region of the Gaussian case of Figure \ref{f3} is
\begin{equation}
\mathcal{C}_{s}^{(Bi)}=\bigcup_{0\leq \alpha\leq 1, 0\leq \beta\leq 1}
\left\{
\begin{array}{ll}
(R_{1}, R_{2}):\\
R_{1}+R_{2}\leq \frac{1}{2}\log(1+\frac{P_{1}+P_{2}+2\sqrt{(1-\alpha)P_{1}P_{2}}}{N_{0}})-\frac{1}{2}\log(1+\frac{\alpha P_{1}}{N_{0}+N_{2}})-\frac{1}{2}\log(1+\frac{\beta P_{2}}{N_{0}+N_{1}})\\
R_{1}+R_{2}\leq \frac{1}{2}\log(1+\frac{P_{1}+P_{2}+2\sqrt{(1-\beta)P_{1}P_{2}}}{N_{0}})-\frac{1}{2}\log(1+\frac{\alpha P_{1}}{N_{0}+N_{2}})-\frac{1}{2}\log(1+\frac{\beta P_{2}}{N_{0}+N_{1}})\\
R_{1}\leq \frac{1}{2}\log(1+\frac{\alpha P_{1}}{N_{0}})-\frac{1}{2}\log(1+\frac{\alpha P_{1}}{N_{0}+N_{2}})\\
R_{2}\leq \frac{1}{2}\log(1+\frac{\beta P_{2}}{N_{0}})-\frac{1}{2}\log(1+\frac{\beta P_{2}}{N_{0}+N_{1}}).
\end{array}
\right\}.
\end{equation}
\end{corollary}
\begin{IEEEproof}
Substituting $R_{e1}=R_{1}$ and $R_{e2}=R_{2}$ into the region $\mathcal{R}^{(Bi)}$ in Theorem \ref{T2},
Corollary 3 is easily obtained.
\end{IEEEproof}

The inner bound on the secrecy capacity $C_{s}^{(Bi)}(R_{2})$ as a function of $R_{2}$ is
\begin{equation}\label{e301xxad}
C_{s}^{(Bi)}(R_{2})=\max_{\alpha}\min
\left\{
\begin{array}{ll}
\frac{1}{2}\log(1+\frac{\alpha P_{1}}{N_{0}})-\frac{1}{2}\log(1+\frac{\alpha P_{1}}{N_{0}+N_{2}}),\\
\frac{1}{2}\log(1+\frac{P_{1}+P_{2}+2\sqrt{(1-\alpha)P_{1}P_{2}}}{N_{0}})-\frac{1}{2}\log(1+\frac{\alpha P_{1}}{N_{0}+N_{2}})-\frac{1}{2}\log(1+\frac{\beta^{*} P_{2}}{N_{0}+N_{1}})-R_{2},\\
\frac{1}{2}\log(1+\frac{P_{1}+P_{2}+2\sqrt{(1-\beta^{*})P_{1}P_{2}}}{N_{0}})-\frac{1}{2}\log(1+\frac{\alpha P_{1}}{N_{0}+N_{2}})-\frac{1}{2}\log(1+\frac{\beta^{*} P_{2}}{N_{0}+N_{1}})-R_{2},
\end{array}
\right\}
\end{equation}
where $\beta^{*}$ is determined by the following equation:
\begin{equation}\label{e301x}
\frac{1}{2}\log(1+\frac{\beta^{*} P_{2}}{N_{0}})-\frac{1}{2}\log(1+\frac{\beta^{*} P_{2}}{N_{0}+N_{1}})=R_{2}.
\end{equation}
\begin{IEEEproof}
The proof of (\ref{e301xxad}) follows directly from Corollary 3.
\end{IEEEproof}

\begin{theorem}\label{T2x}
For the Gaussian case of Figure \ref{f3}, the outer bound $\mathcal{R}^{(Bo)}$ on the capacity-equivocation region $\mathcal{R^{(B)}}$ is given by
\begin{equation}
\mathcal{R}^{(Bo)}=\bigcup_{0\leq \alpha\leq 1, 0\leq \beta\leq 1}
\left\{
\begin{array}{ll}
(R_{1}, R_{2}, R_{e1},R_{e2}):\\
0\leq R_{1}\leq \frac{1}{2}\log(1+\frac{\alpha P_{1}}{N_{0}})\\
0\leq R_{2}\leq \frac{1}{2}\log(1+\frac{\beta P_{2}}{N_{0}})\\
R_{1}+R_{2}\leq \frac{1}{2}\log(1+\frac{P_{1}+P_{2}+2\sqrt{(1-\alpha)P_{1}P_{2}}}{N_{0}})\\
R_{1}+R_{2}\leq \frac{1}{2}\log(1+\frac{P_{1}+P_{2}+2\sqrt{(1-\beta)P_{1}P_{2}}}{N_{0}})\\
R_{e1}\leq R_{1}\\
R_{e2}\leq R_{2}\\
R_{e1}\leq \frac{1}{2}\log(1+\frac{\alpha P_{1}}{N_{0}})-\frac{1}{2}\log(1+\frac{\alpha P_{1}}{N_{0}+N_{2}})\\
R_{e2}\leq \frac{1}{2}\log(1+\frac{\beta P_{2}}{N_{0}})-\frac{1}{2}\log(1+\frac{\beta P_{2}}{N_{0}+N_{1}}).
\end{array}
\right\}.
\end{equation}
\end{theorem}
\begin{IEEEproof}
See Appendix \ref{appen3}.
\end{IEEEproof}

\begin{corollary} The outer bound on the secrecy capacity region of the Gaussian case of Figure \ref{f3} is
\begin{equation}
\mathcal{C}_{s}^{(Bo)}=\bigcup_{0\leq \alpha\leq 1, 0\leq \beta\leq 1}
\left\{
\begin{array}{ll}
(R_{1}, R_{2}):\\
R_{1}+R_{2}\leq \frac{1}{2}\log(1+\frac{P_{1}+P_{2}+2\sqrt{(1-\alpha)P_{1}P_{2}}}{N_{0}})\\
R_{1}+R_{2}\leq \frac{1}{2}\log(1+\frac{P_{1}+P_{2}+2\sqrt{(1-\beta)P_{1}P_{2}}}{N_{0}})\\
R_{1}\leq \frac{1}{2}\log(1+\frac{\alpha P_{1}}{N_{0}})-\frac{1}{2}\log(1+\frac{\alpha P_{1}}{N_{0}+N_{2}})\\
R_{2}\leq \frac{1}{2}\log(1+\frac{\beta P_{2}}{N_{0}})-\frac{1}{2}\log(1+\frac{\beta P_{2}}{N_{0}+N_{1}}).
\end{array}
\right\}.
\end{equation}
\end{corollary}
\begin{IEEEproof}
Substituting $R_{e1}=R_{1}$ and $R_{e2}=R_{2}$ into the region $\mathcal{R}^{(Bo)}$ in Theorem \ref{T2x},
Corollary 4 is easily obtained.
\end{IEEEproof}

The outer bound on the secrecy capacity $C_{s}^{(Bo)}(R_{2})$ as a function of $R_{2}$ is
\begin{equation}\label{e301xx}
C_{s}^{(Bo)}(R_{2})=\max_{\alpha}\min
\left\{
\begin{array}{ll}
\frac{1}{2}\log(1+\frac{\alpha P_{1}}{N_{0}})-\frac{1}{2}\log(1+\frac{\alpha P_{1}}{N_{0}+N_{2}}),\\
\frac{1}{2}\log(1+\frac{P_{1}+P_{2}+2\sqrt{(1-\alpha)P_{1}P_{2}}}{N_{0}})-R_{2},\\
\frac{1}{2}\log(1+\frac{P_{1}+P_{2}+2\sqrt{(1-\beta^{*})P_{1}P_{2}}}{N_{0}})-R_{2},
\end{array}
\right\}
\end{equation}
where $\beta^{*}$ is determined by the following equation:
\begin{equation}\label{e301x}
\frac{1}{2}\log(1+\frac{\beta^{*} P_{2}}{N_{0}})-\frac{1}{2}\log(1+\frac{\beta^{*} P_{2}}{N_{0}+N_{1}})=R_{2}.
\end{equation}
\begin{IEEEproof}
The proof of (\ref{e301xx}) follows directly from Corollary 4.
\end{IEEEproof}

\begin{figure}[htb]
\centering
\includegraphics[scale=0.55]{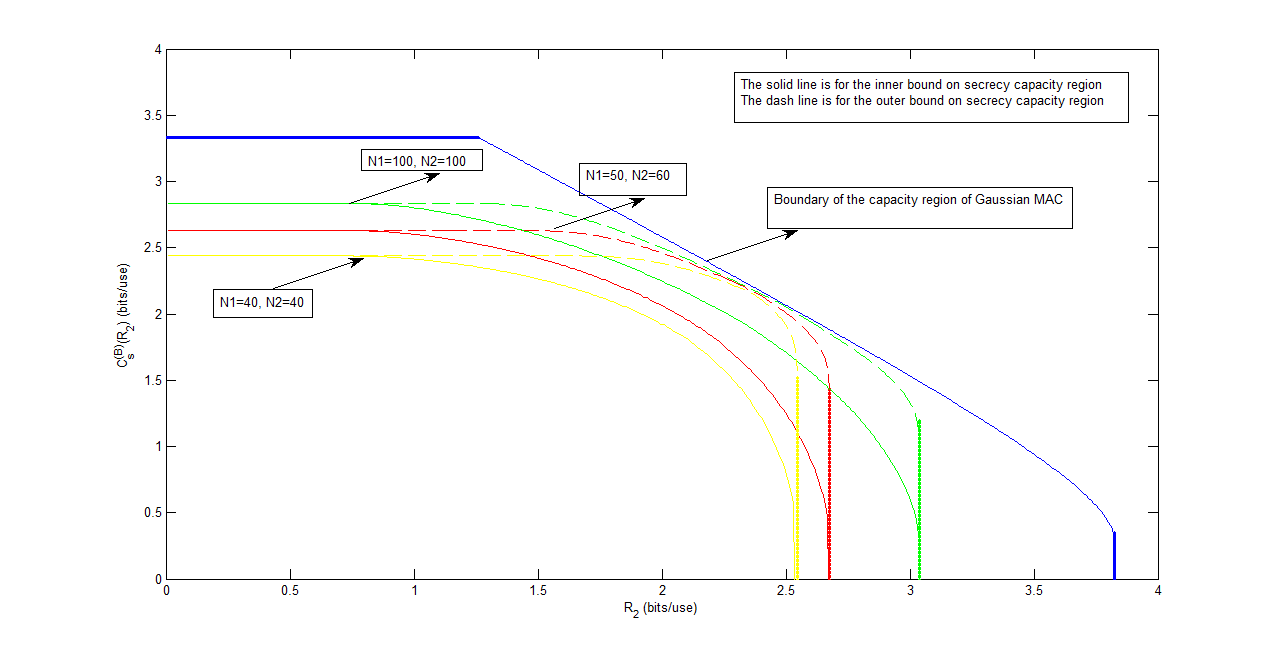}
\caption{Inner and outer bounds on the secrecy capacity regions of the Gaussian case of Figure \ref{f3} and capacity region of corresponding Gaussian MAC,
where $P_{1}=100$, $P_{2}=200$ and $N_{0}=1$.}
\label{f5}
\end{figure}

Figure \ref{f5} plots the inner and outer bounds on the secrecy capacity of the Gaussian case of Figure \ref{f3} for three
values of $N_{1}$ and $N_{2}$. The lines of $C_{s}^{(Bi)}(R_{2})$ and $C_{s}^{(Bo)}(R_{2})$ also serve as the boundaries of the inner and outer bounds on the
secrecy capacity regions if we view
the vertical axis as $R_{1}$. It is easy to see that as $N_{1}$ and $N_{2}$ increase, which implies that the noise level of the wiretap
 channels to both users increases, both users become more confused by the channel outputs. Thus, the inner and outer bounds on the
secrecy capacity region enlarge.

\subsection{The Binary Case of the Model of Figure \ref{f3}\label{sub32}}

In this subsection, we study the following binary case of Figure \ref{f3}. Assume that all channel inputs and outputs take values in
$\{0,1\}$, and the channels are discrete memoryless. The input-output relationship of the channels at each time instant satisfies
\begin{equation}\label{e301xx001}
Y_{i}=X_{1,i}\cdot X_{2,i}, Y_{1,i}=Y_{i}\oplus Z_{1,i}, Y_{2,i}=Y_{i}\oplus Z_{2,i},
\end{equation}
where $1\leq i\leq N$, and $Z^{N}_{1}$, $Z^{N}_{2}$ are composed of $N$ i.i.d. random variables with distributions
$Pr\{Z_{1,i}=1\}=p$ and $Pr\{Z_{2,i}=1\}=q$, respectively. Let $0\leq p, q\leq \frac{1}{2}$.

The following Theorem \ref{T3x} and Theorem \ref{T3} provide inner and outer bounds on the capacity-equivocation region
of the binary MAC with confidential messages.

\begin{theorem}\label{T3}
For the binary case of Figure \ref{f3}, the inner bound on the capacity-equivocation region $\mathcal{R^{(C)}}$ is given by
\begin{equation}
\mathcal{R}^{(Ci)}=\bigcup_{0\leq \alpha\leq 1, 0\leq \beta\leq 1}\left\{
\begin{array}{ll}
(R_{1}, R_{2}, R_{e1},R_{e2}):
\begin{array}{ll}
0\leq R_{1}\leq 1\\
0\leq R_{2}\leq 1\\
R_{1}+R_{2}\leq 1\\
0\leq R_{e1}\leq R_{1}, 0\leq R_{e2}\leq R_{2}\\
R_{e1}\leq h(q)\\
R_{e2}\leq h(p)\\
R_{e1}+R_{e2}\leq h(p)+h(q)-1\\
R_{e1}+R_{2}\leq h(q)\\
R_{e2}+R_{1}\leq h(p)
\end{array}
\end{array}
\right\},
\end{equation}
where $h(p)=-p\log(p)-(1-p)\log(1-p)$, and $h(q)=-q\log(q)-(1-q)\log(1-q)$.
\end{theorem}
\begin{IEEEproof}
See Appendix \ref{appen4}.
\end{IEEEproof}

\begin{corollary} The inner bound on the secrecy capacity region of the binary case of Figure \ref{f3} is
\begin{equation}
\mathcal{C}_{s}^{(Ci)}=\bigcup_{0\leq \alpha\leq 1, 0\leq \beta\leq 1}\left\{
\begin{array}{ll}
(R_{1}, R_{2}):
\begin{array}{ll}
R_{1}+R_{2}\leq h(p)+h(q)-1\\
R_{1}\leq h(q)\\
R_{2}\leq h(p)
\end{array}
\end{array}
\right\}.
\end{equation}
\end{corollary}
\begin{IEEEproof}
Substituting $R_{e1}=R_{1}$ and $R_{e2}=R_{2}$ into the region $\mathcal{R}^{(Ci)}$ in Theorem \ref{T3},
Corollary 5 is easily obtained.
\end{IEEEproof}

\begin{theorem}\label{T3x}
For the binary case of Figure \ref{f3}, the outer bound on the capacity-equivocation region $\mathcal{R^{(C)}}$ is given by
\begin{equation}
\mathcal{R}^{(Co)}=\bigcup_{0\leq \alpha\leq 1, 0\leq \beta\leq 1}\left\{
\begin{array}{ll}
(R_{1}, R_{2}, R_{e1},R_{e2}):
\begin{array}{ll}
0\leq R_{1}\leq 1\\
0\leq R_{2}\leq 1\\
R_{1}+R_{2}\leq 1\\
0\leq R_{e1}\leq R_{1}, 0\leq R_{e2}\leq R_{2}\\
R_{e1}\leq h(q)\\
R_{e2}\leq h(p)
\end{array}
\end{array}
\right\},
\end{equation}
where $h(p)=-p\log(p)-(1-p)\log(1-p)$, and $h(q)=-q\log(q)-(1-q)\log(1-q)$.
\end{theorem}
\begin{IEEEproof}
See Appendix \ref{appen4}.
\end{IEEEproof}

\begin{corollary} The outer bound on the secrecy capacity region of the binary case of Figure \ref{f3} is
\begin{equation}
\mathcal{C}_{s}^{(Co)}=\bigcup_{0\leq \alpha\leq 1, 0\leq \beta\leq 1}\left\{
\begin{array}{ll}
(R_{1}, R_{2}):
\begin{array}{ll}
R_{1}+R_{2}\leq 1\\
R_{1}\leq h(q)\\
R_{2}\leq h(p)
\end{array}
\end{array}
\right\}.
\end{equation}
\end{corollary}
\begin{IEEEproof}
Substituting $R_{e1}=R_{1}$ and $R_{e2}=R_{2}$ into the region $\mathcal{R}^{(Co)}$ in Theorem \ref{T3},
Corollary 6 is easily obtained.
\end{IEEEproof}

\begin{figure}[htb]
\centering
\includegraphics[scale=0.55]{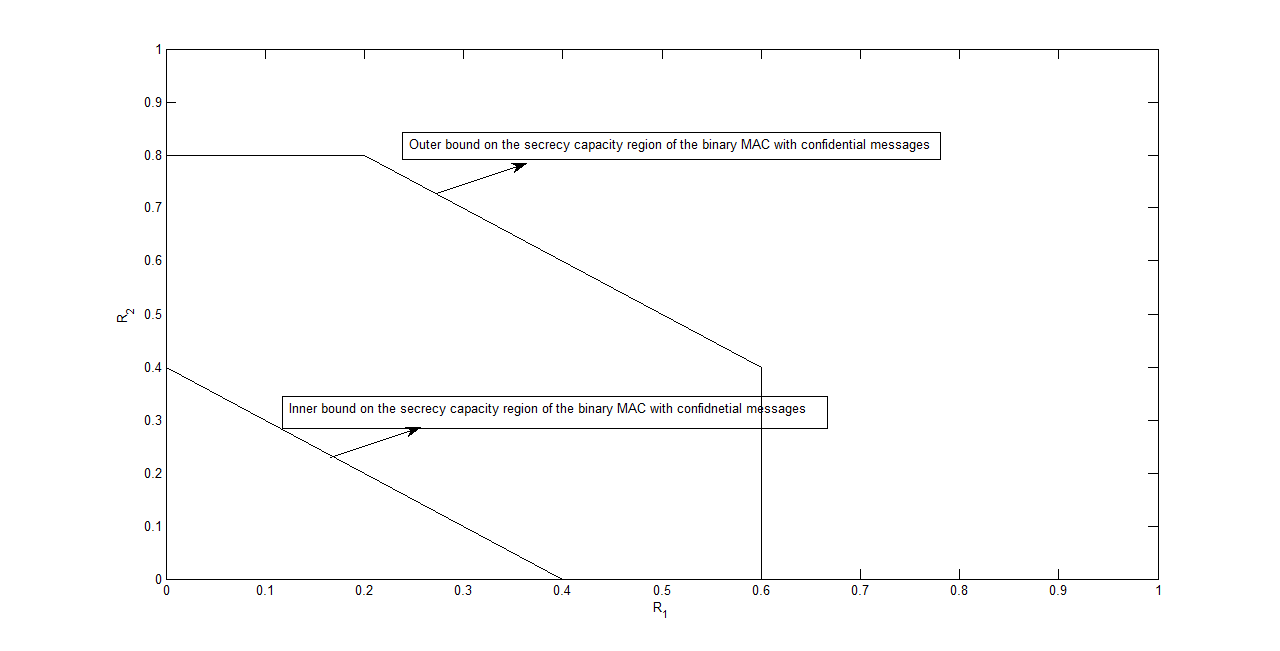}
\caption{Inner and outer bounds on the secrecy capacity region of the binary case of Figure \ref{f3}}
\label{f5r}
\end{figure}

Figure \ref{f5r} plots the inner and outer bounds on the secrecy capacity of the binary case of Figure \ref{f3} for
$h(q)=0.6$ and $h(p)=0.8$. Note that $p$ and $q$ are the cross-over probabilities of the wiretap channels to both users, and as $p$ and $q$ increase,
both users become more and more confused by the channel outputs. When $p=q=\frac{1}{2}$, the inner and outer bounds on the secrecy capacity region
are the same as the capacity region of the MAC.

\section{Multiple Access Wiretap Channel with Cooperative Encoders \label{secV}}
\setcounter{equation}{0}

In this section, a description of the  model of Figure \ref{f4} is given by Definition \ref{def41} to Definition \ref{def53}.
The capacity-equivocation region $\mathcal{R^{(D)}}$ composed of all achievable $(R_{1}, R_{2}, R_{e})$ triples in the model of Figure \ref{f4} is
characterized in Theorem \ref{T4}, where the achievable $(R_{1}, R_{2}, R_{e})$ triple is defined in Definition \ref{def54}.
The capacity-equivocation region of the degraded MAC-WT with cooperative encoders is given in Theorem \ref{T5}.

\begin{definition}(\textbf{Cooperative encoders})\label{def41}
The confidential messages $W_{1}$ and $W_{2}$ take values
in $\mathcal{W}_{1}$ and $\mathcal{W}_{2}$, respectively. $W_{1}$ and $W_{2}$ are independent and uniformly
distributed over their ranges. The inputs of the two encoders are $W_{1}$ and $W_{2}$, while the output of encoder 1
is $X_{1}^{N}$ and the output of encoder 2 is $X_{2}^{N}$.
We assume that the encoders are stochastic encoders, i.e., the encoder
$g_{i}^{N}$ ($i=1,2$) is a matrix of conditional probabilities $g_{i}^{N}(x_{i}^{N}|w_{1},w_{2})$,
where $x_{i}^{N}\in \mathcal{X}_{i}^{N}$, $w_{i}\in \mathcal{W}_{i}$, and $g_{i}^{N}(x_{i}^{N}|w_{1},w_{2})$ is the probability that the messages
$w_{1}$ and $w_{2}$ are encoded as the channel input $x_{i}^{N}$.
\textbf{Note that $X_{1}^{N}$ and $X_{2}^{N}$ are not independent.}

The transmission rates of the confidential messages are $\frac{\log\parallel \mathcal{W}_{1}\parallel}{N}$ and
$\frac{\log\parallel \mathcal{W}_{2}\parallel}{N}$.
\end{definition}

\begin{definition}(\textbf{Channel})\label{def2}
The MAC-WT is a DMC with finite input alphabet
$\mathcal{X}_{1}\times \mathcal{X}_{2}$, finite output alphabet $\mathcal{Y}\times \mathcal{Z}$, and
transition probability $Q_{1}(y,z|x_{1},x_{2})$, where $x_{1}\in \mathcal{X}_{1},x_{2}\in \mathcal{X}_{2},y\in
\mathcal{Y},z\in \mathcal{Z}$. $Q_{1}(y^{N},z^{N}|x_{1}^{N},x_{2}^{N})=\prod_{n=1}^{N}Q_{1}(y_{n},z_{n}|x_{1,n},x_{2,n})$.
The inputs of the channel are $X_{1}^{N}$ and $X_{2}^{N}$, while the outputs are $Y^{N}$ and $Z^{N}$.

The wiretapper's equivocation to the confidential messages $W_{1}$ and $W_{2}$ is defined as
\begin{equation}\label{e501}
\Delta=\frac{1}{N}H(W_{1},W_{2}|Z^{N}).
\end{equation}
\end{definition}

\begin{definition}(\textbf{Decoder})\label{def53}
The decoder is a mapping $f_{D}: \mathcal{Y}^{N}\rightarrow \mathcal{W}_{1}\times \mathcal{W}_{2}$,
with input $Y^{N}$ and outputs $\widehat{W}_{1}$, $\widehat{W}_{2}$. Let $P_{e}$ be the error probability of the receiver
, and it is
defined as $Pr\{(W_{1},W_{2})\neq (\widehat{W}_{1}, \widehat{W}_{2})\}$.
\end{definition}

\begin{definition}(\textbf{Achievable $(R_{1}, R_{2}, R_{e})$ triple in the model of Figure \ref{f4}})\label{def54}
A triple $(R_{1}, R_{2}, R_{e})$ (where $R_{1}, R_{2}, R_{e}>0$) is called
achievable if, for any $\epsilon>0$ (where $\epsilon$ is an arbitrary small positive real number
and $\epsilon\rightarrow 0$), there exists a channel
encoder-decoder $(N, \Delta, P_{e})$ such that
\begin{equation}\label{e503}
\lim_{N\rightarrow \infty}\frac{\log\parallel \mathcal{W}_{1}\parallel}{N}= R_{1} \label{e503a},
\lim_{N\rightarrow \infty}\frac{\log\parallel \mathcal{W}_{2}\parallel}{N}= R_{2} \label{e503b},
\lim_{N\rightarrow \infty}\Delta\geq R_{e} \label{e503c},
P_{e}\leq \epsilon\label{e503d}.
\end{equation}
\end{definition}

\textbf{Theorem \ref{T4} gives a single-letter characterization of the set $\mathcal{R^{(D)}}$, which is composed of all achievable
$(R_{1}, R_{2}, R_{e})$ triples
in the model of Figure \ref{f4}, and it
is proved in Appendix \ref{appen5} and Appendix \ref{appen5a}.}

\begin{theorem}\label{T4}
A single-letter characterization of the region $\mathcal{R^{(D)}}$ is as follows,
\begin{eqnarray*}
&&\mathcal{R^{(D)}}=\{(R_{1}, R_{2}, R_{e}): R_{e}\leq R_{1}+R_{2},\\
&&0\leq R_{1}\leq I(V;Y|U_{2}),\\
&&0\leq R_{2}\leq I(V;Y|U_{1}), \\
&&R_{1}+R_{2}\leq I(V;Y),\\
&&R_{e}\leq I(V;Y|U)-I(V;Z|U)\},
\end{eqnarray*}
where
$(U,U_{1},U_{2})\rightarrow V\rightarrow (X_{1},X_{2})\rightarrow (Y,Z)$,
and $U$, $U_{1}$, $U_{2}$  may be assumed to be (deterministic) functions of $V$.
\end{theorem}

\begin{remark}\label{R4}
There are some notes on Theorem \ref{T4}, see the following.
\begin{itemize}
\item The region $\mathcal{R^{(D)}}$ is convex. The proof is omitted here.

\item The ranges of the random variables $U$, $U_{1}$, $U_{2}$ and $V$ satisfy
$$\|\mathcal{U}\|\leq \|\mathcal{X}_{1}\|\|\mathcal{X}_{2}\|+1,$$
$$\|\mathcal{U}_{1}\|\leq \|\mathcal{X}_{1}\|\|\mathcal{X}_{2}\|,$$
$$\|\mathcal{U}_{2}\|\leq \|\mathcal{X}_{1}\|\|\mathcal{X}_{2}\|,$$
$$\|\mathcal{V}\|\leq (\|\mathcal{X}_{1}\|\|\mathcal{X}_{2}\|+1)^{2}\|\mathcal{X}_{1}\|^{2}\|\mathcal{X}_{2}\|^{2}.$$
The proof is in Appendix \ref{appen7}.

\item The secrecy capacity region $\mathcal{C}_{s}^{(D)}$ is the set of pairs $(R_{1},R_{2})$ such that
$(R_{1}, R_{2}, R_{e}=R_{1}+R_{2})\in \mathcal{R^{(D)}}$.
\begin{corollary}
\begin{eqnarray*}
&&\mathcal{C}_{s}^{(D)}=\{(R_{1}, R_{2}): R_{1}\leq I(V;Y|U_{2}), \\
&&R_{2}\leq I(V;Y|U_{1}),\\
&&R_{1}+R_{2}\leq I(V;Y|U)-I(V;Z|U)\}.
\end{eqnarray*}
\end{corollary}
\begin{IEEEproof}
Corollary 7 is easy to be checked by substituting $R_{e}=R_{1}+R_{2}$ into $\mathcal{R^{(D)}}$.
\end{IEEEproof}

\end{itemize}
\end{remark}

\textbf{For the degraded MAC-WT with cooperative encoders, i.e., $(W_{1},W_{2})\rightarrow (X_{1},X_{2})\rightarrow Y\rightarrow Z$,
the capacity-equivocation region $\mathcal{R^{(E)}}$ is given in the following Theorem \ref{T5},
and it is proved in Appendix \ref{appen8}.}

\begin{theorem}\label{T5}
A single-letter characterization of the region $\mathcal{R^{(E)}}$ for the degraded MAC-WT with cooperative encoders is as follows,
\begin{eqnarray*}
&&\mathcal{R^{(E)}}=\{(R_{1}, R_{2}, R_{e}): R_{e}\leq R_{1}+R_{2},\\
&&0\leq R_{1}\leq I(V;Y|U_{2}),\\
&&0\leq R_{2}\leq I(V;Y|U_{1}), \\
&&R_{1}+R_{2}\leq I(V;Y),\\
&&R_{e}\leq I(V;Y)-I(V;Z)\},
\end{eqnarray*}
where
$(U_{1},U_{2})\rightarrow V\rightarrow (X_{1},X_{2})\rightarrow (Y,Z)$,
and $U_{1}$, $U_{2}$  may be assumed to be (deterministic) functions of $V$.
\end{theorem}

\begin{remark}\label{R5}
There are some notes on Theorem \ref{T5}, see the following.
\begin{itemize}

\item For the degraded case, the last bound in Theorem \ref{T5} can be obtained from the corresponding bound of Theorem \ref{T4}, see the following.
\begin{eqnarray}\label{e50aa}
R_{e}&\leq& I(V;Y|U)-I(V;Z|U)\nonumber\\
&=&I(V;Y)-I(V;Z)-(I(U;Y)-I(U;Z))\nonumber\\
&\leq&I(V;Y)-I(V;Z).
\end{eqnarray}
Therefore, the converse proof of Theorem \ref{T5} is directly obtained from that of Theorem \ref{T4} and
the above (\ref{e50aa}).

\item The region $\mathcal{R^{(E)}}$ is convex. The proof is omitted here.

\item The ranges of the random variables $U_{1}$, $U_{2}$ and $V$ satisfy
$$\|\mathcal{U}_{1}\|\leq \|\mathcal{X}_{1}\|\|\mathcal{X}_{2}\|,$$
$$\|\mathcal{U}_{2}\|\leq \|\mathcal{X}_{1}\|\|\mathcal{X}_{2}\|,$$
$$\|\mathcal{V}\|\leq (\|\mathcal{X}_{1}\|\|\mathcal{X}_{2}\|+1)\|\mathcal{X}_{1}\|^{2}\|\mathcal{X}_{2}\|^{2}.$$
The proof is similar to Appendix \ref{appen7}, and it is omitted here.

\item The secrecy capacity region $\mathcal{C}_{s}^{(E)}$ is the set of pairs $(R_{1},R_{2})$ such that
$(R_{1}, R_{2}, R_{e}=R_{1}+R_{2})\in \mathcal{R^{(E)}}$.
\begin{corollary}
\begin{eqnarray*}
&&\mathcal{C}_{s}^{(E)}=\{(R_{1}, R_{2}): R_{1}\leq I(V;Y|U_{2}), \\
&&R_{2}\leq I(V;Y|U_{1}),\\
&&R_{1}+R_{2}\leq I(V;Y)-I(V;Z)\}.
\end{eqnarray*}
\end{corollary}
\begin{IEEEproof}
Corollary 8 is easy to be checked by substituting $R_{e}=R_{1}+R_{2}$ into $\mathcal{R^{(E)}}$
\end{IEEEproof}

\end{itemize}
\end{remark}

\section{Degraded Multiple Access Wiretap Channel with Non-Cooperative Encoders\label{secVI}}
\setcounter{equation}{0}

In this section, we will present inner and outer bounds on the capacity-equivocation of the degraded MAC-WT with non-cooperative encoders.
For the non-cooperative model, the input of encoder 1 is $W_{1}$, while the output of encoder 1 is $X_{1}^{N}$.
Similarly, the input and output of encoder 2 are $W_{2}$ and $X_{2}^{N}$, respectively. The encoders are stochastic encoders, i.e.,
the encoder $g_{i}^{*N}$ ($i=1,2$) is a matrix of conditional probabilities $g_{i}^{*N}(x_{i}^{N}|w_{i})$,
where $x_{i}^{N}\in \mathcal{X}_{i}^{N}$, $w_{i}\in \mathcal{W}_{i}$, and $g_{i}^{*N}(x_{i}^{N}|w_{i})$ is the probability that the message
$w_{i}$ is encoded as the channel input $x_{i}^{N}$.
Note that $X_{1}^{N}$ and $X_{2}^{N}$ are independent.

\textbf{The inner and outer bounds on the capacity-equivocation region $\mathcal{R^{(F)}}$ of the degraded MAC-WT with non-cooperative encoders are provided
in Theorem \ref{T6} and Theorem \ref{T7}, respectively, and they
are proved in Appendix \ref{appen9} and Appendix \ref{appen10}.}

\begin{theorem}\label{T6}
(\textbf{Inner bound}) A single-letter characterization of the region $\mathcal{R^{(G)}}$ is as follows,
\begin{eqnarray*}
\mathcal{R^{(G)}}=\mathcal{L}^{(1)}\bigcup \mathcal{L}^{(2)}\bigcup \mathcal{L}^{(3)}\bigcup \mathcal{L}^{(4)},
\end{eqnarray*}
where
\begin{eqnarray*}
\mathcal{L}^{(1)}=\left\{
\begin{array}{ll}
(R_{1}, R_{2},R_{e}):\\
R_{1}\leq I(X_{1};Y)-I(X_{1};Z)\\
R_{2}\leq I(X_{2};Y|X_{1})\\
R_{1}+R_{2}\leq I(X_{1},X_{2};Y)\\
R_{e}\leq R_{1}+R_{2}\\
R_{e}\leq I(X_{2};Y|X_{1})-I(X_{2};Z|X_{1})+R_{1}
\end{array}
\right\},
\end{eqnarray*}

\begin{eqnarray*}
\mathcal{L}^{(2)}=
\left\{
\begin{array}{ll}
(R_{1}, R_{2},R_{e}):\\
R_{1}\leq I(X_{1};Y|X_{2})\\
R_{2}\leq I(X_{2};Y)-I(X_{2};Z)\\
R_{1}+R_{2}\leq I(X_{1},X_{2};Y)\\
R_{e}\leq R_{1}+R_{2}\\
R_{e}\leq I(X_{1};Y|X_{2})-I(X_{1};Z|X_{2})+R_{2}
\end{array}
\right\},
\end{eqnarray*}

\begin{eqnarray*}
\mathcal{L}^{(3)}=
\left\{
\begin{array}{ll}
(R_{1}, R_{2},R_{e}):\\
I(X_{1};Y)-I(X_{1};Z)\leq R_{1}\leq I(X_{1};Y)\\
I(X_{2};Y|X_{1})-I(X_{2};Z|X_{1})\leq R_{2}\leq I(X_{2};Y|X_{1})\\
R_{1}+R_{2}\leq I(X_{1},X_{2};Y)\\
R_{e}\leq R_{1}+R_{2}\\
R_{e}\leq I(X_{1},X_{2};Y)-I(X_{1},X_{2};Z)
\end{array}
\right\},
\end{eqnarray*}

\begin{eqnarray*}
\mathcal{L}^{(4)}=
\left\{
\begin{array}{ll}
(R_{1}, R_{2},R_{e}):\\
I(X_{1};Y|X_{2})-I(X_{1};Z|X_{2})\leq R_{1}\leq I(X_{1};Y|X_{2})\\
I(X_{2};Y)-I(X_{2};Z)\leq R_{2}\leq I(X_{2};Y)\\
R_{1}+R_{2}\leq I(X_{1},X_{2};Y)\\
R_{e}\leq R_{1}+R_{2}\\
R_{e}\leq I(X_{1},X_{2};Y)-I(X_{1},X_{2};Z)
\end{array}
\right\},
\end{eqnarray*}
and $X_{1}$, $X_{2}$, $Y$ and $Z$ satisfy $(X_{1},X_{2})\rightarrow Y\rightarrow Z$, and $\mathcal{R^{(G)}}\subseteq \mathcal{R^{(F)}}$.
\end{theorem}

\begin{remark}\label{R6}
There are some notes on Theorem \ref{T6}, see the following.
\begin{itemize}

\item The region $\mathcal{R^{(G)}}$ is convex. The proof is omitted here.

\item The secrecy capacity region $\mathcal{C}_{s}^{(F)}$ is the set of pairs $(R_{1},R_{2})$ such that
$(R_{1}, R_{2}, R_{e}=R_{1}+R_{2})\in \mathcal{R^{(F)}}$.
\begin{corollary}
(\textbf{Inner bound on secrecy capacity region}) The secrecy capacity region $\mathcal{C}_{s}^{(F)}$ satisfies
$\mathcal{C}_{s}^{(G)}\subseteq \mathcal{C}_{s}^{(F)}$, where
\begin{eqnarray*}
\mathcal{C}_{s}^{(G)}=\left\{
\begin{array}{ll}
(R_{1}, R_{2}):\\
R_{1}\leq I(X_{1};Y)-I(X_{1};Z)\\
R_{2}\leq I(X_{2};Y|X_{1})-I(X_{2};Z|X_{1})
\end{array}
\right\}\bigcup
\left\{
\begin{array}{ll}
(R_{1}, R_{2}):\\
R_{1}\leq I(X_{1};Y|X_{2})-I(X_{1};Z|X_{2})\\
R_{2}\leq I(X_{2};Y)-I(X_{2};Z)
\end{array}
\right\}.
\end{eqnarray*}
\end{corollary}
\begin{IEEEproof}
Corollary 9 is easy to be checked by substituting $R_{e}=R_{1}+R_{2}$ into $\mathcal{R^{(G)}}$.
\end{IEEEproof}

\end{itemize}
\end{remark}

\begin{theorem}\label{T7}
(\textbf{Outer bound}) A single-letter characterization of the region $\mathcal{R^{(H)}}$ is as follows,
\begin{eqnarray*}
\mathcal{R^{(H)}}=\left\{
\begin{array}{ll}
(R_{1}, R_{2},R_{e}):\\
R_{1}\leq I(X_{1};Y|X_{2})\\
R_{2}\leq I(X_{2};Y|X_{1})\\
R_{1}+R_{2}\leq I(X_{1},X_{2};Y)\\
R_{e}\leq R_{1}+R_{2}\\
R_{e}\leq I(X_{1},X_{2};Y)-I(X_{1},X_{2};Z)
\end{array}
\right\},
\end{eqnarray*}
where $X_{1}$, $X_{2}$, $Y$ and $Z$ satisfy $(X_{1},X_{2})\rightarrow Y\rightarrow Z$, and $\mathcal{R^{(F)}}\subseteq \mathcal{R^{(H)}}$.
\end{theorem}

\begin{remark}\label{R7}
There are some notes on Theorem \ref{T7}, see the following.
\begin{itemize}

\item The region $\mathcal{R^{(H)}}$ is convex. The proof is omitted here.

\item
\begin{corollary}
(\textbf{Outer bound on secrecy capacity region}) The secrecy capacity region $\mathcal{C}_{s}^{(F)}$ satisfies $\mathcal{C}_{s}^{(F)}\subseteq \mathcal{C}_{s}^{(H)}$, where
\begin{eqnarray*}
\mathcal{C}_{s}^{(H)}=\left\{
\begin{array}{ll}
(R_{1}, R_{2}):\\
R_{1}\leq I(X_{1};Y|X_{2})\\
R_{2}\leq I(X_{2};Y|X_{1})\\
R_{1}+R_{2}\leq I(X_{1},X_{2};Y)-I(X_{1},X_{2};Z)
\end{array}
\right\}.
\end{eqnarray*}
\end{corollary}
\begin{IEEEproof}
Corollary 10 is easy to be checked by substituting $R_{e}=R_{1}+R_{2}$ into $\mathcal{R^{(H)}}$.
\end{IEEEproof}

\end{itemize}
\end{remark}

\begin{figure}[htb]
\centering
\includegraphics[scale=0.7]{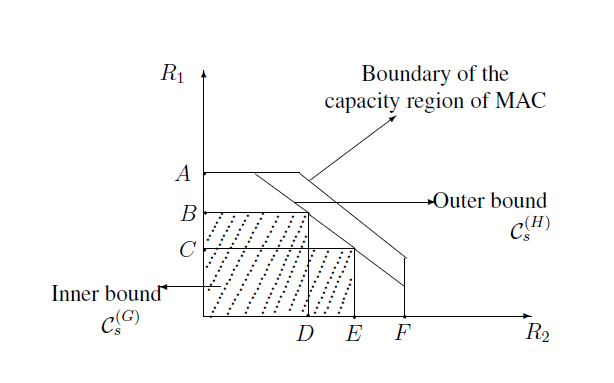}
\caption{The inner and outer bounds on the secrecy capacity region $\mathcal{C}_{s}^{(F)}$, and the capacity region of the MAC,
where $A=I(X_{1};Y|X_{2})$, $B=I(X_{1};Y)-I(X_{1};Z)$, $C=I(X_{1};Y|X_{2})-I(X_{1};Z|X_{2})$,
$D=I(X_{2};Y|X_{1})$, $E=I(X_{2};Y)-I(X_{2};Z)$ and $F=I(X_{2};Y|X_{1})-I(X_{2};Z|X_{1})$.}
\label{jsacaa}
\end{figure}

To understand the relationship of the inner bound $\mathcal{C}_{s}^{(G)}$, the outer bound $\mathcal{C}_{s}^{(H)}$
and the capacity region of the MAC, we plot Figure \ref{jsacaa} for illustration.

\section{Binary Degraded MAC-WT with Cooperative (or Non-Cooperative) Encoders\label{secV}}
\setcounter{equation}{0}

\subsection{The Binary Case of the Degraded MAC-WT with Cooperative Encoders\label{sub32}}

In this subsection, we study the binary case of the degraded MAC-WT with cooperative encoders. Assume that all channel inputs and outputs take values in
$\{0,1\}$, and the channels are discrete memoryless. The input-output relationship of the channels at each time instant satisfies
\begin{equation}\label{e8a}
Y_{i}=X_{1,i}\cdot X_{2,i}, \ \ Z_{i}=Y_{i}\oplus Z^{*}_{i},
\end{equation}
where $1\leq i\leq N$, and $Z^{*N}$ is composed of $N$ i.i.d. random variables with distribution
$Pr\{Z^{*}_{i}=1\}=p$ and $Pr\{Z^{*}_{i}=0\}=1-p$. Let $0\leq p\leq \frac{1}{2}$.

\begin{theorem}\label{T8}
For the binary case of the degraded MAC-WT with cooperative encoders, the capacity-equivocation region $\mathcal{R^{(I)}}$ is given by
\begin{equation}
\mathcal{R^{(I)}}=\left\{
\begin{array}{ll}
(R_{1}, R_{2}, R_{e}):
\begin{array}{ll}
0\leq R_{1}\leq 1\\
0\leq R_{2}\leq 1\\
R_{1}+R_{2}\leq 1\\
R_{e}\leq R_{1}+R_{2}\\
R_{e}\leq h(p)
\end{array}
\end{array}
\right\},
\end{equation}
where $h(p)=-p\log(p)-(1-p)\log(1-p)$.
\end{theorem}
\begin{IEEEproof}
By calculating the mutual information terms in Theorem \ref{T5}, Theorem \ref{T8} is easy to be checked, and therefore, the proof is omitted here.
\end{IEEEproof}

\begin{corollary} The secrecy capacity region of the binary case of the degraded MAC-WT with cooperative encoders is
\begin{equation}
\mathcal{C}_{s}^{(I)}=\left\{
\begin{array}{ll}
(R_{1}, R_{2}):
\begin{array}{ll}
R_{1}+R_{2}\leq h(p)\\
R_{1}\leq 1\\
R_{2}\leq 1
\end{array}
\end{array}
\right\}.
\end{equation}
\end{corollary}
\begin{IEEEproof}
Substituting $R_{e}=R_{1}+R_{2}$ into the region $\mathcal{R^{(I)}}$ in Theorem \ref{T8},
Corollary 11 is easily obtained.
\end{IEEEproof}

\begin{figure}[htb]
\centering
\includegraphics[scale=0.7]{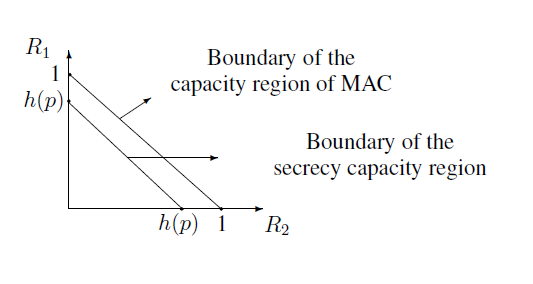}
\caption{The secrecy capacity region of the binary case of the degraded MAC-WT with cooperative encoders, and
the capacity region of the binary MAC}
\label{jsacb}
\end{figure}

Figure \ref{jsacb} shows the secrecy capacity region of the binary case of the degraded MAC-WT with cooperative encoders,
and the capacity region of the binary MAC. It is easy to see that as $p\rightarrow \frac{1}{2}$, the secrecy capacity region tends to be the
capacity region of the binary MAC.

\subsection{The Binary Case of the Degraded MAC-WT with Non-Cooperative Encoders\label{sub32}}

In this subsection, we study the binary case of the degraded MAC-WT with non-cooperative encoders. Assume that all channel inputs and outputs take values in
$\{0,1\}$, and the channels are discrete memoryless. The input-output relationship of the channels at each time instant satisfies
\begin{equation}\label{e9a}
Y_{i}=X_{1,i}\cdot X_{2,i}, \ \ Z_{i}=Y_{i}\oplus Z^{*}_{i},
\end{equation}
where $1\leq i\leq N$, and $Z^{*N}$ is composed of $N$ i.i.d. random variables with distribution
$Pr\{Z^{*}_{i}=1\}=p$ and $Pr\{Z^{*}_{i}=0\}=1-p$. Let $0\leq p\leq \frac{1}{2}$.

\begin{theorem}\label{T9}
For the binary case of the degraded MAC-WT with non-cooperative encoders, the inner bound on
the secrecy capacity region is coincident with the corresponding outer bound. Therefore,
the secrecy capacity region $\mathcal{R^{(J)}}$ is
\begin{equation}
\mathcal{R^{(J)}}=\left\{
\begin{array}{ll}
(R_{1}, R_{2}):
\begin{array}{ll}
0\leq R_{1}\leq h(p)\\
0\leq R_{2}\leq h(p)\\
R_{1}+R_{2}\leq h(p)
\end{array}
\end{array}
\right\},
\end{equation}
where $h(p)=-p\log(p)-(1-p)\log(1-p)$.
\end{theorem}
\begin{IEEEproof}
See Appendix \ref{appen11}.
\end{IEEEproof}

\begin{figure}[htb]
\centering
\includegraphics[scale=0.7]{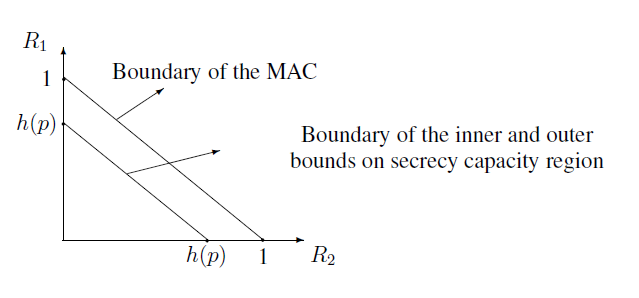}
\caption{The secrecy capacity region of the binary case of the degraded MAC-WT with non-cooperative encoders, and
the capacity region of the binary MAC}
\label{jsacc}
\end{figure}

Figure \ref{jsacc} shows the secrecy capacity region of the binary case of the degraded MAC-WT with non-cooperative encoders,
and the capacity region of the binary MAC. It is easy to see that as $p\rightarrow \frac{1}{2}$, the secrecy capacity region tends to be the
capacity region of the binary MAC.

\section{Conclusion\label{s8}}

In this paper, first, we study the model of degraded MAC with confidential messages. The inner and outer bounds on the capacity-equivocation region
and the secrecy capacity region are provided for this model. Second, as two examples, the binary and Gaussian cases of
the degraded MAC with confidential messages are studied, and the inner and outer bounds on the capacity-equivocation regions are also given
 for the two examples.

Third, we investigate the MAC-WT with cooperative encoders. The capacity-equivocation regions and the corresponding
secrecy capacity regions are determined for both the general model and the degraded model. Fourth, for the model of degraded MAC-WT
with non-cooperative encoders, we present inner and outer bounds on the capacity-equivocation region. Finally, we give binary examples
for the degraded MAC-WT with cooperative (or non-cooperative) encoders.

\renewcommand{\theequation}{\arabic{equation}}
\appendices\section{Proof of Theorem \ref{T1}\label{appen1b}}

Suppose
$(R_{1},R_{2},R_{e1},R_{e2})\in \mathcal{R}^{Ai}$, we will show that $(R_{1},R_{2},R_{e1},R_{e2})$ is achievable.
Without loss of generality, the proof of Theorem \ref{T1} is considered into the following four cases.

\begin{itemize}
\item \textbf{(Case 1)} If
$I(X_{2};Y)\geq I(X_{2};Y_{1}|X_{1})$ and $R_{2}\leq I(X_{2};Y)$, we only need to prove that the tuple $(R_{1},R_{2},R_{e1},R_{e2})$ satisfying
$R_{e1}=I(X_{1};Y|X_{2})-I(X_{1};Y_{2}|X_{2})$ and $R_{e2}=I(X_{2};Y)-I(X_{2};Y_{1}|X_{1})$, is achievable.

\item \textbf{(Case 2)} If
$I(X_{2};Y)\geq I(X_{2};Y_{1}|X_{1})$ and $R_{2}\geq I(X_{2};Y)$,
we only need to prove that the tuple $(R_{1},R_{2},R_{e1},R_{e2})$ satisfying
$R_{e1}=I(X_{1};Y|X_{2})-I(X_{1};Y_{2}|X_{2})+I(X_{2};Y)-R_{2}$ and $R_{e2}=I(X_{2};Y)-I(X_{2};Y_{1}|X_{1})$ is achievable.

\item \textbf{(Case 3)} If
$I(X_{2};Y)\leq I(X_{2};Y_{1}|X_{1})$ and $R_{2}\leq I(X_{2};Y_{1}|X_{1})$,
we only need to prove that the tuple $(R_{1},R_{2},R_{e1},R_{e2})$ satisfying
$R_{e1}=I(X_{1};Y|X_{2})-I(X_{1};Y_{2}|X_{2})+I(X_{2};Y)-I(X_{2};Y_{1}|X_{1})$ and $R_{e2}=0$ is achievable.

\item \textbf{(Case 4)} If
$I(X_{2};Y)\leq I(X_{2};Y_{1}|X_{1})$ and $R_{2}\geq I(X_{2};Y_{1}|X_{1})$,
we only need to prove that the tuple $(R_{1},R_{2},R_{e1},R_{e2})$ satisfying
$R_{e1}=I(X_{1};Y|X_{2})-I(X_{1};Y_{2}|X_{2})+I(X_{2};Y)-R_{2}$ and $R_{e2}=0$ is achievable.

\end{itemize}

Now the
remainder of this section is organized as follows. Some preliminaries about typical sequences are introduced in Subsection \ref{app2.1}.
For the four cases, the construction of the code is
introduced in Subsection \ref{app2.2}. For any given $\epsilon> 0$, the
proofs of $\lim_{N\rightarrow \infty}\frac{\log\parallel \mathcal{W}_{1}\parallel}{N}= R_{1}$,
$\lim_{N\rightarrow \infty}\frac{\log\parallel \mathcal{W}_{2}\parallel}{N}= R_{2}$,
$\lim_{N\rightarrow \infty}\Delta_{1}\geq R_{e1}$, $\lim_{N\rightarrow \infty}\Delta_{2}\geq R_{e2}$
and $P_{e}\leq \epsilon$ are given in Subsection \ref{app2.3}.

\subsection{Preliminaries\label{app2.1}}

\begin{itemize}
\item Given a probability mass function $p_{V}(v)$, for any $\eta>0$, let $T^{N}_{V}(\eta)$ be the strong typical set
of all $v^{N}$ such that $|p_{V}(v)-\frac{c_{v^{N}}(v)}{N}|<\eta$ for all $v\in \mathcal{V}$, where
$c_{v^{N}}(v)$ is the number of occurences of the letter $v$ in the $v^{N}$. We say that
the sequences $v^{N}\in T^{N}_{V}(\eta)$ are \textbf{$V$-typical}.

\item Analogously, given a joint probability mass function $p_{VW}(v,w)$,
for any $\eta>0$, let $T^{N}_{VW}(\eta)$ be the joint strong typical set of all pairs $(v^{N},w^{N})$ such that $|p_{VW}(v,w)-\frac{c_{v^{N},w^{N}}(v,w)}{N}|<\eta$
for all $v\in \mathcal{V}$ and $w\in \mathcal{W}$,
where $c_{v^{N},w^{N}}(v,w)$ is the number of occurences of $(v,w)$ in the pair of sequences $(v^{N},w^{N})$.
We say that the pairs of sequences $(v^{N},w^{N})\in T^{N}_{VW}(\eta)$ are \textbf{$VW$-typical}.

\item Moreover, $w^{N}$ is called
\textbf{$W|V$-generated} by $v^{N}$ iff $v^{N}$ is $V$- typical and $(v^{N},w^{N})\in T^{N}_{VW}(\eta)$. For any given $v^{N}\in T^{N}_{V}(\eta)$,
define $T^{N}_{W|V}(\eta)=\{w^{N}:w^{N}\ \mbox{is} \ W|V\mbox{-generated} \ \mbox{by} \ v^{N}\}$.

\item

\begin{lemma}\label{appL1} For any $v^{N}\in T^{N}_{V}(\eta)$,
$$2^{-N(H(V)+\eta^{*})}\leq p_{V^{N}}(v^{N})\leq 2^{-N(H(V)-\eta^{*})},$$
where $\eta^{*}\rightarrow 0$ as $\eta\rightarrow 0$.
\end{lemma}

\end{itemize}

\subsection{ Coding Construction\label{app2.2}}

\textbf{The code constructions for the four cases are almost the same (by using Wyner's random binning technique), except that the total number of $x_{1}^{N}$ and $x_{2}^{N}$ are different,
see the followings.}

\begin{itemize}
\item For case 1, the existence of the encoder-decoder is under the sufficient conditions that
$R_{e1}=I(X_{1};Y|X_{2})-I(X_{1};Y_{2}|X_{2})$ and $R_{e2}=I(X_{2};Y)-I(X_{2};Y_{1}|X_{1})$.
Given a tuple $(R_{1},R_{2},R_{e1},R_{e2})$, choose a joint probability mass function $p_{X_{1},X_{2},Y,Y_{1},Y_{2}}(x_{1},x_{2},y,y_{1},y_{2})$
such that
$$0\leq R_{1}\leq I(X_{1};Y|X_{2}),\ \ 0\leq R_{2}\leq I(X_{2};Y|X_{1}),\ \ R_{1}+R_{2}\leq I(X_{1},X_{2};Y),$$
$$R_{e1}\leq R_{1}, \ \ R_{e1}=I(X_{1};Y|X_{2})-I(X_{1};Y_{2}|X_{2}),$$
$$R_{e2}\leq R_{2}, \ \ R_{e2}=I(X_{2};Y)-I(X_{2};Y_{1}|X_{1}).$$
It is easy to check that the last three inequalities in Theorem \ref{T1}
hold by using the conditions that $R_{e1}=I(X_{1};Y|X_{2})-I(X_{1};Y_{2}|X_{2})$ and $R_{e2}=I(X_{2};Y)-I(X_{2};Y_{1}|X_{1})$.

The confidential message sets $\mathcal{W}_{1}$ and $\mathcal{W}_{2}$ satisfy the following conditions:
\begin{equation}\label{appe1}
\lim_{N\rightarrow \infty}\frac{1}{N}\log\parallel \mathcal{W}_{1}\parallel=R_{1}, \ \
\lim_{N\rightarrow \infty}\frac{1}{N}\log\parallel \mathcal{W}_{2}\parallel=R_{2}.
\end{equation}

Code-book generation for case 1:
\begin{itemize}

\item Generate $2^{N(I(X_{1};Y|X_{2})-\epsilon_{N})}$ codewords $x_{1}^{N}$ ($\epsilon_{N}\rightarrow \infty$ as $N\rightarrow \infty$),
and each of them is uniformly drawn from the strong typical set
$T^{N}_{X_{1}}(\eta)$. Divide the $2^{N(I(X_{1};Y|X_{2})-\epsilon_{N})}$ codewords into $2^{NR_{1}}$ bins, and each bin corresponds to
a specific value in $\mathcal{W}_{1}$.

\item Analogously, generate $2^{N(I(X_{2};Y)-\epsilon_{N})}$ codewords $x_{2}^{N}$, and each of them is uniformly drawn from the strong typical set
$T^{N}_{X_{2}}(\eta)$. Divide the $2^{N(I(X_{2};Y)-\epsilon_{N})}$ codewords into $2^{NR_{2}}$ bins, and each bin corresponds to
a specific value in $\mathcal{W}_{2}$.

\end{itemize}

\item For case 2, the existence of the encoder-decoder is under the sufficient conditions that
$R_{e1}=I(X_{1};Y|X_{2})-I(X_{1};Y_{2}|X_{2})+I(X_{2};Y)-R_{2}$ and $R_{e2}=I(X_{2};Y)-I(X_{2};Y_{1}|X_{1})$.
Given a tuple $(R_{1},R_{2},R_{e1},R_{e2})$, choose a joint probability mass function $p_{X_{1},X_{2},Y,Y_{1},Y_{2}}(x_{1},x_{2},y,y_{1},y_{2})$
such that
$$0\leq R_{1}\leq I(X_{1};Y|X_{2}),\ \ 0\leq R_{2}\leq I(X_{2};Y|X_{1}),\ \ R_{1}+R_{2}\leq I(X_{1},X_{2};Y),$$
$$R_{e1}\leq R_{1}, \ \ R_{e1}=I(X_{1};Y|X_{2})-I(X_{1};Y_{2}|X_{2})+I(X_{2};Y)-R_{2},$$
$$R_{e2}\leq R_{2}, \ \ R_{e2}=I(X_{2};Y)-I(X_{2};Y_{1}|X_{1}).$$
It is easy to check that the last three inequalities in Theorem \ref{T1}
hold by using the conditions that $R_{e1}=I(X_{1};Y|X_{2})-I(X_{1};Y_{2}|X_{2})+I(X_{2};Y)-R_{2}$ and $R_{e2}=I(X_{2};Y)-I(X_{2};Y_{1}|X_{1})$.

The confidential message sets $\mathcal{W}_{1}$ and $\mathcal{W}_{2}$ also satisfy (\ref{appe1}).

Code-book generation for case 2:
\begin{itemize}

\item Generate $2^{N(I(X_{1};Y|X_{2})+I(X_{2};Y)-R_{2}-\epsilon_{N})}$ codewords $x_{1}^{N}$ ($\epsilon_{N}\rightarrow \infty$ as $N\rightarrow \infty$),
and each of them is uniformly drawn from the strong typical set
$T^{N}_{X_{1}}(\eta)$. Divide the $2^{N(I(X_{1};Y|X_{2})+I(X_{2};Y)-R_{2}-\epsilon_{N})}$ codewords into $2^{NR_{1}}$ bins, and each bin corresponds to
a specific value in $\mathcal{W}_{1}$.

\item Generate $2^{N(I(X_{2};Y)-\epsilon_{N})}$ codewords $x_{2}^{N}$, and each of them is uniformly drawn from the strong typical set
$T^{N}_{X_{2}}(\eta)$. Divide the $2^{N(I(X_{2};Y)-\epsilon_{N})}$ codewords into $2^{NR_{2}}$ bins, and each bin corresponds to
a specific value in $\mathcal{W}_{2}$.

\end{itemize}

\item For case 3, the existence of the encoder-decoder is under the sufficient conditions that
$R_{e1}=I(X_{1};Y|X_{2})-I(X_{1};Y_{2}|X_{2})+I(X_{2};Y)-I(X_{2};Y_{1}|X_{1})$ and $R_{e2}=0$.
Given a tuple $(R_{1},R_{2},R_{e1},R_{e2})$, choose a joint probability mass function $p_{X_{1},X_{2},Y,Y_{1},Y_{2}}(x_{1},x_{2},y,y_{1},y_{2})$
such that
$$0\leq R_{1}\leq I(X_{1};Y|X_{2}),\ \ 0\leq R_{2}\leq I(X_{2};Y|X_{1}),\ \ R_{1}+R_{2}\leq I(X_{1},X_{2};Y),$$
$$R_{e1}\leq R_{1}, \ \ R_{e1}=I(X_{1};Y|X_{2})-I(X_{1};Y_{2}|X_{2})+I(X_{2};Y)-I(X_{2};Y_{1}|X_{1}),$$
$$R_{e2}\leq R_{2}, \ \ R_{e2}=0.$$
It is easy to check that the last three inequalities in Theorem \ref{T1}
hold by using the conditions that $R_{e1}=I(X_{1};Y|X_{2})-I(X_{1};Y_{2}|X_{2})+I(X_{2};Y)-I(X_{2};Y_{1}|X_{1})$ and $R_{e2}=0$.

The confidential message sets $\mathcal{W}_{1}$ and $\mathcal{W}_{2}$ also satisfy (\ref{appe1}).

Code-book generation for case 3:
\begin{itemize}

\item Generate $2^{N(I(X_{1};Y|X_{2})+I(X_{2};Y)-I(X_{2};Y_{1}|X_{1})-\epsilon_{N})}$ codewords $x_{1}^{N}$ ($\epsilon_{N}\rightarrow \infty$ as $N\rightarrow \infty$),
and each of them is uniformly drawn from the strong typical set
$T^{N}_{X_{1}}(\eta)$. Divide the $2^{N(I(X_{1};Y|X_{2})+I(X_{2};Y)-I(X_{2};Y_{1}|X_{1})-\epsilon_{N})}$ codewords into $2^{NR_{1}}$ bins, and each bin corresponds to
a specific value in $\mathcal{W}_{1}$.

\item Generate $2^{N(I(X_{2};Y)-\epsilon_{N})}$ codewords $x_{2}^{N}$, and each of them is uniformly drawn from the strong typical set
$T^{N}_{X_{2}}(\eta)$. Divide the $2^{N(I(X_{2};Y)-\epsilon_{N})}$ codewords into $2^{NR_{2}}$ bins, and each bin corresponds to
a specific value in $\mathcal{W}_{2}$.

\end{itemize}

\item For case 4, the existence of the encoder-decoder is under the sufficient conditions that
$R_{e1}=I(X_{1};Y|X_{2})-I(X_{1};Y_{2}|X_{2})+I(X_{2};Y)-R_{2}$ and $R_{e2}=0$.
Given a tuple $(R_{1},R_{2},R_{e1},R_{e2})$, choose a joint probability mass function $p_{X_{1},X_{2},Y,Y_{1},Y_{2}}(x_{1},x_{2},y,y_{1},y_{2})$
such that
$$0\leq R_{1}\leq I(X_{1};Y|X_{2}),\ \ 0\leq R_{2}\leq I(X_{2};Y|X_{1}),\ \ R_{1}+R_{2}\leq I(X_{1},X_{2};Y),$$
$$R_{e1}\leq R_{1}, \ \ R_{e1}=I(X_{1};Y|X_{2})-I(X_{1};Y_{2}|X_{2})+I(X_{2};Y)-R_{2},$$
$$R_{e2}\leq R_{2}, \ \ R_{e2}=0.$$
It is easy to check that the last three inequalities in Theorem \ref{T1}
hold by using the conditions that $R_{e1}=I(X_{1};Y|X_{2})-I(X_{1};Y_{2}|X_{2})+I(X_{2};Y)-R_{2}$ and $R_{e2}=0$.

The confidential message sets $\mathcal{W}_{1}$ and $\mathcal{W}_{2}$ also satisfy (\ref{appe1}).

Code-book generation for case 4:
\begin{itemize}

\item Generate $2^{N(I(X_{1};Y|X_{2})+I(X_{2};Y)-R_{2}-\epsilon_{N})}$ codewords $x_{1}^{N}$ ($\epsilon_{N}\rightarrow \infty$ as $N\rightarrow \infty$),
and each of them is uniformly drawn from the strong typical set
$T^{N}_{X_{1}}(\eta)$. Divide the $2^{N(I(X_{1};Y|X_{2})+I(X_{2};Y)-R_{2}-\epsilon_{N})}$ codewords into $2^{NR_{1}}$ bins, and each bin corresponds to
a specific value in $\mathcal{W}_{1}$.

\item Generate $2^{N(I(X_{2};Y)-\epsilon_{N})}$ codewords $x_{2}^{N}$, and each of them is uniformly drawn from the strong typical set
$T^{N}_{X_{2}}(\eta)$. Divide the $2^{N(I(X_{2};Y)-\epsilon_{N})}$ codewords into $2^{NR_{2}}$ bins, and each bin corresponds to
a specific value in $\mathcal{W}_{2}$.

\end{itemize}

\item \textbf{(Decoding scheme for all cases)} For a given $y^{N}$, try to find a pair of sequences
$(x_{1}^{N}(\hat{w}_{1}),x_{2}^{N}(\hat{w}_{2}))$ such that
$(x_{1}^{N}(\hat{w}_{1}),x_{2}^{N}(\hat{w}_{2}),y^{N})\in T^{N}_{X_{1}X_{2}Y}(\epsilon)$.
If there exist sequences with the same indices
 $\hat{w}_{1}$ and $\hat{w}_{2}$, put out the corresponding
$\hat{w}_{1}$ and $\hat{w}_{2}$, else declare a decoding error.

\end{itemize}

\subsection{Proof of the Achievability\label{app2.3}}

By using the above definitions, it is easy to verify that  $\lim_{N\rightarrow \infty}\frac{\log\parallel \mathcal{W}_{1}\parallel}{N}= R_{1}$ and
$\lim_{N\rightarrow \infty}\frac{\log\parallel \mathcal{W}_{2}\parallel}{N}= R_{2}$ for the two cases.

From the standard techniques as in \cite[Ch. 14]{cover}, we have $P_{e}\leq \epsilon$ for all cases.

It remains to show that $\lim_{N\rightarrow \infty}\Delta_{1}\geq R_{e1}$ and $\lim_{N\rightarrow \infty}\Delta_{2}\geq R_{e2}$
for the four cases, see the followings.

\begin{itemize}

\item \textbf{(Proof of  $\lim_{N\rightarrow \infty}\Delta_{1}\geq R_{e1}$ and $\lim_{N\rightarrow \infty}\Delta_{2}\geq R_{e2}$ for case 1)}

First, we compute the following equivocation rate of $W_{1}$.
\begin{eqnarray}\label{appe4}
\lim_{N\rightarrow \infty}\Delta_{1}&\triangleq&\lim_{N\rightarrow \infty}\frac{1}{N}H(W_{1}|Y_{2}^{N},X_{2}^{N})\nonumber\\
&=&\lim_{N\rightarrow \infty}\frac{1}{N}(H(W_{1},Y_{2}^{N},X_{2}^{N})-H(Y_{2}^{N},X_{2}^{N}))\nonumber\\
&=&\lim_{N\rightarrow \infty}\frac{1}{N}(H(W_{1},Y_{2}^{N},X_{1}^{N},X_{2}^{N})-H(X_{1}^{N}|W_{1},Y_{2}^{N},X_{2}^{N})-H(Y_{2}^{N},X_{2}^{N}))\nonumber\\
&\stackrel{(a)}=&\lim_{N\rightarrow \infty}\frac{1}{N}(H(Y_{2}^{N}|X_{1}^{N},X_{2}^{N})+H(W_{1},X_{1}^{N})+H(X_{2}^{N})-H(X_{1}^{N}|W_{1},Y_{2}^{N},X_{2}^{N})\nonumber\\
&&-H(Y_{2}^{N},X_{2}^{N}))\nonumber\\
&=&\lim_{N\rightarrow \infty}\frac{1}{N}(H(Y_{2}^{N}|X_{1}^{N},X_{2}^{N})+H(X_{1}^{N}|W_{1})+H(W_{1})-H(X_{1}^{N}|W_{1},Y_{2}^{N},X_{2}^{N})\nonumber\\
&&-H(Y_{2}^{N}|X_{2}^{N}))\nonumber\\
&=&\lim_{N\rightarrow \infty}\frac{1}{N}(H(X_{1}^{N}|W_{1})+H(W_{1})-H(X_{1}^{N}|W_{1},Y_{2}^{N},X_{2}^{N})-I(Y_{2}^{N};X_{1}^{N}|X_{2}^{N}),
\end{eqnarray}
where (a) is from $W_{1}\rightarrow (X_{1}^{N},X_{2}^{N})\rightarrow Y_{2}^{N}$ and the fact that $X_{2}^{N}$ is independent of
$W_{1}$ and $X_{1}^{N}$.

The first term in (\ref{appe4}) can be bounded as follows.
\begin{equation}\label{appe5}
\lim_{N\rightarrow \infty}\frac{1}{N}H(X_{1}^{N}|W_{1})\geq I(X_{1};Y|X_{2})-R_{1},
\end{equation}
where (\ref{appe5}) is from the property of the strong typical sequences.

The second term in (\ref{appe4}) is as follows.
\begin{equation}\label{appe6}
\lim_{N\rightarrow \infty}\frac{1}{N}H(W_{1})=R_{1}.
\end{equation}

For the third term in (\ref{appe4}), we have
\begin{equation}\label{appe7}
\lim_{N\rightarrow \infty}\frac{1}{N}H(X_{1}^{N}|W_{1},Y_{2}^{N},X_{2}^{N})=0.
\end{equation}
This is because for a given $w_{1}$, there are $2^{N(I(X_{1};Y|X_{2})-\epsilon_{N}-R_{1})}$ codewords left for $x_{1}^{N}$.
Then note that
\begin{eqnarray*}
I(X_{1};Y|X_{2})-\epsilon_{N}-R_{1}&\leq&I(X_{1};Y|X_{2})-\epsilon_{N}-R_{e1}\nonumber\\
&=&I(X_{1};Y|X_{2})-\epsilon_{N}-(I(X_{1};Y|X_{2})-I(X_{1};Y_{2}|X_{2}))\nonumber\\
&=&I(X_{1};Y_{2}|X_{2})-\epsilon_{N},
\end{eqnarray*}
and $\epsilon_{N}\rightarrow 0$ as $N\rightarrow \infty$. From the standard channel coding theorem and the Fano's inequality,
we have (\ref{appe7}).

For the fourth term in (\ref{appe4}), we have
\begin{equation}\label{appe8}
\lim_{N\rightarrow \infty}\frac{1}{N}I(Y_{2}^{N};X_{1}^{N}|X_{2}^{N})\leq I(X_{1};Y_{2}|X_{2}),
\end{equation}
and this is from a standard technique as in \cite[p. 343]{CK}.

Substituting (\ref{appe5}), (\ref{appe6}), (\ref{appe7}) and (\ref{appe8}) into (\ref{appe4}), we have
\begin{equation}\label{appe9}
\lim_{N\rightarrow \infty}\Delta_{1}\geq I(X_{1};Y|X_{2})-I(X_{1};Y_{2}|X_{2})=R_{e1}.
\end{equation}
$\lim_{N\rightarrow \infty}\Delta_{1}\geq R_{e1}$ is proved.
Analogously, we can prove that $\lim_{N\rightarrow \infty}\Delta_{2}\geq R_{e2}$, see the following.
\begin{eqnarray}\label{appe4rr}
\lim_{N\rightarrow \infty}\Delta_{2}&\triangleq&\lim_{N\rightarrow \infty}\frac{1}{N}H(W_{2}|Y_{1}^{N},X_{1}^{N})\nonumber\\
&=&\lim_{N\rightarrow \infty}\frac{1}{N}(H(W_{2},Y_{1}^{N},X_{1}^{N})-H(Y_{1}^{N},X_{1}^{N}))\nonumber\\
&=&\lim_{N\rightarrow \infty}\frac{1}{N}(H(W_{2},Y_{1}^{N},X_{2}^{N},X_{1}^{N})-H(X_{2}^{N}|W_{2},Y_{1}^{N},X_{1}^{N})-H(Y_{1}^{N},X_{1}^{N}))\nonumber\\
&\stackrel{(a)}=&\lim_{N\rightarrow \infty}\frac{1}{N}(H(Y_{1}^{N}|X_{2}^{N},X_{1}^{N})+H(W_{2},X_{2}^{N})+H(X_{1}^{N})-H(X_{2}^{N}|W_{2},Y_{1}^{N},X_{1}^{N})\nonumber\\
&&-H(Y_{1}^{N},X_{1}^{N}))\nonumber\\
&=&\lim_{N\rightarrow \infty}\frac{1}{N}(H(Y_{1}^{N}|X_{2}^{N},X_{1}^{N})+H(X_{2}^{N}|W_{2})+H(W_{2})-H(X_{2}^{N}|W_{2},Y_{1}^{N},X_{1}^{N})\nonumber\\
&&-H(Y_{1}^{N}|X_{1}^{N}))\nonumber\\
&=&\lim_{N\rightarrow \infty}\frac{1}{N}(H(X_{2}^{N}|W_{2})+H(W_{2})-H(X_{2}^{N}|W_{2},Y_{1}^{N},X_{1}^{N})-I(Y_{1}^{N};X_{2}^{N}|X_{1}^{N}),
\end{eqnarray}
where (a) is from $W_{2}\rightarrow (X_{1}^{N},X_{2}^{N})\rightarrow Y_{1}^{N}$ and the fact that $X_{1}^{N}$ is independent of
$W_{2}$ and $X_{2}^{N}$.

The first term in (\ref{appe4rr}) can be bounded as follows.
\begin{equation}\label{appe5rr}
\lim_{N\rightarrow \infty}\frac{1}{N}H(X_{2}^{N}|W_{2})\geq I(X_{2};Y)-R_{2},
\end{equation}
where (\ref{appe5rr}) is from the property of the strong typical sequences.

The second term in (\ref{appe4rr}) is as follows.
\begin{equation}\label{appe6rr}
\lim_{N\rightarrow \infty}\frac{1}{N}H(W_{2})=R_{2}.
\end{equation}

For the third term in (\ref{appe4rr}), we have
\begin{equation}\label{appe7rr}
\lim_{N\rightarrow \infty}\frac{1}{N}H(X_{2}^{N}|W_{2},Y_{1}^{N},X_{1}^{N})=0.
\end{equation}
This is because for a given $w_{2}$, there are $2^{N(I(X_{2};Y)-\epsilon_{N}-R_{2})}$ codewords left for $x_{2}^{N}$.
Then note that
\begin{eqnarray*}
I(X_{2};Y)-\epsilon_{N}-R_{2}&\leq&I(X_{2};Y)-\epsilon_{N}-R_{e2}\nonumber\\
&=&I(X_{2};Y)-\epsilon_{N}-(I(X_{2};Y)-I(X_{2};Y_{1}|X_{1}))\nonumber\\
&=&I(X_{2};Y_{1}|X_{1})-\epsilon_{N},
\end{eqnarray*}
and $\epsilon_{N}\rightarrow 0$ as $N\rightarrow \infty$. From the standard channel coding theorem and the Fano's inequality,
we have (\ref{appe7rr}).

For the fourth term in (\ref{appe4rr}), we have
\begin{equation}\label{appe8rr}
\lim_{N\rightarrow \infty}\frac{1}{N}I(Y_{1}^{N};X_{2}^{N}|X_{1}^{N})\leq I(X_{2};Y_{1}|X_{1}),
\end{equation}
and this is from a standard technique as in \cite[p. 343]{CK}.

Substituting (\ref{appe5rr}), (\ref{appe6rr}), (\ref{appe7rr}) and (\ref{appe8rr}) into (\ref{appe4rr}), we have
\begin{equation}\label{appe9rr}
\lim_{N\rightarrow \infty}\Delta_{2}\geq I(X_{2};Y)-I(X_{2};Y_{1}|X_{1})=R_{e2}.
\end{equation}
$\lim_{N\rightarrow \infty}\Delta_{2}\geq R_{e2}$ is proved.
Therefore, the proof for case 1 is completed.

\item \textbf{(Proof of  $\lim_{N\rightarrow \infty}\Delta_{1}\geq R_{e1}$ and $\lim_{N\rightarrow \infty}\Delta_{2}\geq R_{e2}$ for case 2, case 3 and case 4)}
Note that (\ref{appe4}) and (\ref{appe4rr}) also hold for case 2, case 3 and case 4, and the proofs of $\lim_{N\rightarrow \infty}\Delta_{1}\geq R_{e1}$
and $\lim_{N\rightarrow \infty}\Delta_{2}\geq R_{e2}$
are similar to
that of case 1. Therefore, we omit the proof here.
The proof for case 2, case 3 and case 4 is completed.

\end{itemize}

The proof of  Theorem \ref{T1} is completed.

\renewcommand{\theequation}{\arabic{equation}}
\section{Proof of Theorem \ref{T1x}\label{appen1}}
\setcounter{equation}{0}

In this section, we prove Theorem \ref{T1x}: all the achievable $(R_{1}, R_{2}, R_{e1}, R_{e2})$ tuples are
contained in the set $\mathcal{R}^{(Ao)}$, i.e., for any achievable tuple, there exist random variables $X_{1}$, $X_{2}$,
$Y$, $Y_{1}$ and $Y_{2}$ such that the inequalities in Theorem \ref{T1x} hold, and $(X_{1},X_{2})\rightarrow Y\rightarrow (Y_{1},Y_{2})$
forms a Markov chain. We will prove the inequalities of Theorem \ref{T1x} in the remainder of this section.

\textbf{(Proof of $0\leq R_{1}\leq I(X_{1};Y|X_{2})$)} The proof of this inequality is as follows.
\begin{eqnarray}\label{app1}
\frac{1}{N}H(W_{1})&\stackrel{(a)}\leq&\frac{1}{N}(I(W_{1};Y^{N})+\delta(P_{e}))\nonumber\\
&\stackrel{(b)}\leq&\frac{1}{N}(I(X^{N}_{1};Y^{N})+\delta(P_{e}))\nonumber\\
&\stackrel{(c)}\leq&\frac{1}{N}(I(X^{N}_{1};Y^{N}|X^{N}_{2})+\delta(P_{e}))\nonumber\\
&\stackrel{(d)}\leq&\frac{1}{N}(\sum_{i=1}^{N}I(X_{1,i};Y_{i}|X_{2,i})+\delta(P_{e}))\nonumber\\
&\stackrel{(e)}=&I(X_{1};Y|X_{2})+\frac{\delta(P_{e})}{N},
\end{eqnarray}
where (a) is from the Fano's inequality, (b) is from the data processing theorem,
(c) is from the fact that $X^{N}_{1}$ and $X^{N}_{2}$ are independent, (d) is from the discrete memoryless property of the channel,
and (e) is from the definitions that $X_{1}\triangleq (X_{1,J}, J)$, $X_{2}\triangleq (X_{2,J}, J)$, $Y\triangleq Y_{J}$,
where $J$ is a random variable (uniformly distributed over $\{1,2,...,N\}$), and it is independent of $X_{1,i}$, $X_{2,i}$ and $Y_{i}$.

By using $P_{e}\leq \epsilon$, $R_{1}=\lim_{N\rightarrow \infty}\frac{H(W_{1})}{N}$ and (\ref{app1}), it is easy to see that
$0\leq R_{1}\leq I(X_{1};Y|X_{2})$.

\textbf{(Proof of $0\leq R_{2}\leq I(X_{2};Y|X_{1})$)} The proof is similar to the proof of $0\leq R_{1}\leq I(X_{1};Y|X_{2})$, and it is omitted here.

\textbf{(Proof of $0\leq R_{1}+R_{2}\leq I(X_{1}, X_{2};Y)$)}
\begin{eqnarray}\label{app2}
\frac{1}{N}H(W_{1},W_{2})&\stackrel{(1)}\leq&\frac{1}{N}(I(W_{1},W_{2};Y^{N})+\delta(P_{e}))\nonumber\\
&\stackrel{(2)}\leq&\frac{1}{N}(I(X^{N}_{1},X^{N}_{2};Y^{N})+\delta(P_{e}))\nonumber\\
&\stackrel{(3)}\leq&\frac{1}{N}(\sum_{i=1}^{N}I(X_{1,i},X_{2,i};Y_{i})+\delta(P_{e}))\nonumber\\
&\stackrel{(4)}=&I(X_{1},X_{2};Y)+\frac{\delta(P_{e})}{N},
\end{eqnarray}
where (1) is from the Fano's inequality, (2) is from the data processing theorem, (3) is from the discrete memoryless property of the channel,
and (4) is from the definitions that $X_{1}\triangleq (X_{1,J}, J)$, $X_{2}\triangleq (X_{2,J}, J)$, $Y\triangleq Y_{J}$.

By using $P_{e}\leq \epsilon$, $R_{1}=\lim_{N\rightarrow \infty}\frac{H(W_{1})}{N}$, $R_{2}=\lim_{N\rightarrow \infty}\frac{H(W_{2})}{N}$
and (\ref{app2}), it is easy to see that
$0\leq R_{1}+R_{2}\leq I(X_{1}, X_{2};Y)$.

\textbf{(Proof of $0\leq R_{e1}\leq R_{1}$ and $0\leq R_{e2}\leq R_{2}$)} The two inequalities are obtained by the following equations.
\begin{equation}\label{app3}
R_{e1}\leq\lim_{N\rightarrow \infty}\frac{1}{N}H(W_{1}|Y_{2}^{N},X_{2}^{N})\leq \lim_{N\rightarrow \infty}\frac{1}{N}H(W_{1})=R_{1}.
\end{equation}
\begin{equation}\label{app4}
R_{e2}\leq\lim_{N\rightarrow \infty}\frac{1}{N}H(W_{2}|Y_{1}^{N},X_{1}^{N})\leq \lim_{N\rightarrow \infty}\frac{1}{N}H(W_{2})=R_{2}.
\end{equation}

\textbf{(Proof of $R_{e1}\leq I(X_{1};Y|X_{2})-I(X_{1};Y_{2}|X_{2})$)} The proof is obtained by the following (\ref{app5}), (\ref{app6}),
and $P_{e}\leq \epsilon$.
\begin{eqnarray}\label{app5}
R_{e1}&\leq&\lim_{N\rightarrow \infty}\frac{1}{N}H(W_{1}|Y_{2}^{N},X_{2}^{N})\nonumber\\
&\stackrel{(a)}\leq&\lim_{N\rightarrow \infty}\frac{1}{N}(H(W_{1}|Y_{2}^{N},X_{2}^{N})+\delta(P_{e})-H(W_{1}|Y_{2}^{N},X_{2}^{N},Y^{N}))\nonumber\\
&=&\lim_{N\rightarrow \infty}\frac{1}{N}(I(W_{1};Y^{N}|Y_{2}^{N},X_{2}^{N})+\delta(P_{e}))\nonumber\\
&\leq&\lim_{N\rightarrow \infty}\frac{1}{N}(H(Y^{N}|Y_{2}^{N},X_{2}^{N})-H(Y^{N}|Y_{2}^{N},X_{2}^{N},W_{1},X_{1}^{N})+\delta(P_{e}))\nonumber\\
&\stackrel{(b)}=&\lim_{N\rightarrow \infty}\frac{1}{N}(H(Y^{N}|Y_{2}^{N},X_{2}^{N})-H(Y^{N}|Y_{2}^{N},X_{2}^{N},X_{1}^{N})+\delta(P_{e}))\nonumber\\
&=&\lim_{N\rightarrow \infty}\frac{1}{N}(I(Y^{N};X_{1}^{N}|Y_{2}^{N},X_{2}^{N})+\delta(P_{e})),
\end{eqnarray}
where (a) is from the Fano's inequality, and (b) is from $W_{1}\rightarrow (Y_{2}^{N},X_{2}^{N},X_{1}^{N})\rightarrow Y^{N}$.

\begin{eqnarray}\label{app6}
I(Y^{N};X_{1}^{N}|Y_{2}^{N},X_{2}^{N})&=&H(X_{1}^{N}|Y_{2}^{N},X_{2}^{N})-H(X_{1}^{N}|Y_{2}^{N},X_{2}^{N},Y^{N})\nonumber\\
&\stackrel{(c)}=&H(X_{1}^{N}|Y_{2}^{N},X_{2}^{N})-H(X_{1}^{N}|X_{2}^{N},Y^{N})\nonumber\\
&=&H(X_{1}^{N}|Y_{2}^{N},X_{2}^{N})-H(X_{1}^{N}|X_{2}^{N},Y^{N})-H(X_{1}^{N}|X_{2}^{N})+H(X_{1}^{N}|X_{2}^{N})\nonumber\\
&=&I(X_{1}^{N};Y^{N}|X_{2}^{N})-I(X_{1}^{N};Y_{2}^{N}|X_{2}^{N})\nonumber\\
&=&H(Y^{N}|X_{2}^{N})-H(Y^{N}|X_{1}^{N},X_{2}^{N})-H(Y_{2}^{N}|X_{2}^{N})+H(Y_{2}^{N}|X_{1}^{N},X_{2}^{N})\nonumber\\
&=&\sum_{i=1}^{N}(H(Y_{i}|Y^{i-1},X_{2}^{N})-H(Y_{i}|X_{1,i},X_{2,i})-H(Y_{2,i}|Y_{2}^{i-1},X_{2}^{N})+H(Y_{2,i}|X_{1,i},X_{2,i}))\nonumber\\
&\leq&\sum_{i=1}^{N}(H(Y_{i}|Y^{i-1},X_{2}^{N})-H(Y_{i}|X_{1,i},X_{2,i})-H(Y_{2,i}|Y_{2}^{i-1},X_{2}^{N},Y^{i-1})+H(Y_{2,i}|X_{1,i},X_{2,i}))\nonumber\\
&\stackrel{(d)}=&\sum_{i=1}^{N}(H(Y_{i}|Y^{i-1},X_{2}^{N})-H(Y_{i}|X_{1,i},X_{2,i})-H(Y_{2,i}|X_{2}^{N},Y^{i-1})+H(Y_{2,i}|X_{1,i},X_{2,i}))\nonumber\\
&\stackrel{(e)}=&\sum_{i=1}^{N}(H(Y_{i}|U_{i},X_{2,i})-H(Y_{i}|X_{1,i},X_{2,i})-H(Y_{2,i}|U_{i},X_{2,i})+H(Y_{2,i}|X_{1,i},X_{2,i}))\nonumber\\
&=&\sum_{i=1}^{N}(H(Y_{i}|U_{i},X_{2,i})-H(Y_{i}|X_{1,i},X_{2,i})-H(Y_{i}|X_{2,i})+H(Y_{i}|X_{2,i})\nonumber\\
&&-H(Y_{2,i}|U_{i},X_{2,i})+H(Y_{2,i}|X_{1,i},X_{2,i})+H(Y_{2,i}|X_{2,i})-H(Y_{2,i}|X_{2,i}))\nonumber\\
&=&\sum_{i=1}^{N}(I(X_{1,i};Y_{i}|X_{2,i})-I(U_{i};Y_{i}|X_{2,i})+I(U_{i};Y_{2,i}|X_{2,i})-I(X_{1,i};Y_{2,i}|X_{2,i}))\nonumber\\
&=&\sum_{i=1}^{N}(I(X_{1,i};Y_{i}|X_{2,i})-I(X_{1,i};Y_{2,i}|X_{2,i})+H(U_{i}|X_{2,i},Y_{i})-H(U_{i}|X_{2,i},Y_{2,i}))\nonumber\\
&\stackrel{(f)}=&\sum_{i=1}^{N}(I(X_{1,i};Y_{i}|X_{2,i})-I(X_{1,i};Y_{2,i}|X_{2,i})+H(U_{i}|X_{2,i},Y_{i},Y_{2,i})-H(U_{i}|X_{2,i},Y_{2,i}))\nonumber\\
&=&\sum_{i=1}^{N}(I(X_{1,i};Y_{i}|X_{2,i})-I(X_{1,i};Y_{2,i}|X_{2,i})-I(U_{i};Y_{i}|X_{2,i},Y_{2,i}))\nonumber\\
&\leq&\sum_{i=1}^{N}(I(X_{1,i};Y_{i}|X_{2,i})-I(X_{1,i};Y_{2,i}|X_{2,i}))\nonumber\\
&\stackrel{(g)}=&I(X_{1};Y|X_{2})-I(X_{1};Y_{2}|X_{2}),
\end{eqnarray}
where (c) is from $Y_{2}^{N}\rightarrow (X_{2}^{N},Y^{N})\rightarrow X_{1}^{N}$,
(d) is from $Y_{2}^{i-1}\rightarrow (X_{2}^{N},Y^{i-1})\rightarrow Y_{2,i}$,
(e) is from the definition $U_{i}\triangleq (Y^{i-1},X_{2,i+1}^{N},X_{2}^{i-1})$,
(f) is from $U_{i}\rightarrow (X_{2,i},Y_{i})\rightarrow Y_{2,i}$,
and (g) is from the definitions that $X_{1}\triangleq (X_{1,J}, J)$, $X_{2}\triangleq (X_{2,J}, J)$, $Y\triangleq Y_{J}$, $Y_{2}\triangleq Y_{2,J}$
where $J$ is a random variable (uniformly distributed over $\{1,2,...,N\}$), and it is independent of $X_{1,i}$, $X_{2,i}$, $Y_{i}$ and $Y_{2,i}$.

\textbf{(Proof of $R_{e2}\leq I(X_{2};Y|X_{1})-I(X_{2};Y_{1}|X_{1})$)} The proof is analogous to the proof of
$R_{e1}\leq I(X_{1};Y|X_{2})-I(X_{1};Y_{2}|X_{2})$.

The Markov chain $(X_{1},X_{2})\rightarrow Y\rightarrow (Y_{1},Y_{2})$ is directly obtained from the definitions
$X_{1}\triangleq (X_{1,J}, J)$, $X_{2}\triangleq (X_{2,J}, J)$, $Y\triangleq Y_{J}$, $Y_{2}\triangleq Y_{2,J}$ and $Y_{1}\triangleq Y_{1,J}$.

The proof of Theorem \ref{T1x} is completed.

\renewcommand{\theequation}{\arabic{equation}}
\section{Proof of Theorem \ref{T2} and Theorem \ref{T2x}\label{appen3}}

\subsection{Proof of Theorem \ref{T2}\label{appen3.1}}
The achievability proof follows by computing the mutual information terms in Theorem \ref{T1} with the following joint distributions:

$$X_{1}^{'}\sim \mathcal{N}(0, \alpha P_{1}) \ \mbox{and} \ X_{2}^{'}\sim \mathcal{N}(0, \beta P_{2}).$$

$$X_{1}=\sqrt{\frac{(1-\alpha)P_{1}}{P_{2}}}X_{2}+X^{'}_{1} \ \mbox{and} \ X_{2}=\sqrt{\frac{(1-\beta)P_{2}}{P_{1}}}X_{1}+X^{'}_{2}.$$

$X_{1}^{'}$ is independent of $X_{2}^{'}$.

\subsection{Proof of Theorem \ref{T2x}\label{appen3.2}}

The proof of $R_{e1}\leq \frac{1}{2}\log(1+\frac{\alpha P_{1}}{N_{0}})-\frac{1}{2}\log(1+\frac{\alpha P_{1}}{N_{0}+N_{2}})$
and $R_{e2}\leq \frac{1}{2}\log(1+\frac{\beta P_{2}}{N_{0}})-\frac{1}{2}\log(1+\frac{\beta P_{2}}{N_{0}+N_{1}})$
can be directly obtained from \cite[p. 1000-1001]{LP} by letting $R_{0}=0$ and $Q$ be a constant.
Other inequalities in Theorem \ref{T2} are from the capacity region of the Gaussian MAC, and they are easily obtained
from \cite[p. 999-1000]{LP}. Therefore, the full details are omitted here.

\renewcommand{\theequation}{\arabic{equation}}
\section{Proof of Theorem \ref{T3} and Theorem \ref{T3x}\label{appen4}}

The proof of Theorem \ref{T3} is along the lines of Appendix \ref{appen1b}.
The proof of Theorem \ref{T3x} is obtained by computing the mutual information terms in Theorem \ref{T1x}, see the followings.

All the random variables take values in $\{0, 1\}$.
Let $Pr\{X_{1}=0\}=\alpha$, $Pr\{X_{1}=1\}=1-\alpha$, $Pr\{X_{2}=0\}=\beta$ and $Pr\{X_{2}=1\}=1-\beta$.
Note that $X_{1}$, $X_{2}$, $Y$, $Y_{1}$ and $Y_{2}$ satisfy
\begin{equation}\label{appeee1}
Y=X_{1}\cdot X_{2}, Y_{1}=Y\oplus Z_{1}, Y_{2}=Y\oplus Z_{2},
\end{equation}
where $X_{1}$ is independent of $X_{2}$, and $Pr\{Z_{1}=0\}=1-p$, $Pr\{Z_{1}=1\}=p$, $Pr\{Z_{2}=0\}=1-q$, $Pr\{Z_{2}=1\}=q$.

The joint probability $p_{X_{1}X_{2}Y}$ is calculated by (\ref{appeee2}).
\begin{equation}\label{appeee2}
p_{X_{1},X_{2},Y}(x_{1},x_{2},y)=p_{Y|X_{1},X_{2}}(y|x_{1},x_{2})p_{X_{1}}(x_{1})p_{X_{2}}(x_{2}).
\end{equation}
The joint probability $p_{X_{1}X_{2}Y_{1}}$ is calculated by (\ref{appeee3}).
\begin{equation}\label{appeee3}
p_{X_{1},X_{2},Y_{1}}(x_{1},x_{2},y_{1})=\sum_{y}p_{Y_{1}|Y}(y_{1}|y)p_{Y|X_{1},X_{2}}(y|x_{1},x_{2})p_{X_{1}}(x_{1})p_{X_{2}}(x_{2}).
\end{equation}
The joint probability $p_{X_{1}X_{2}Y_{2}}$ is calculated by (\ref{appeee4}).
\begin{equation}\label{appeee4}
p_{X_{1},X_{2},Y_{2}}(x_{1},x_{2},y_{2})=\sum_{y}p_{Y_{2}|Y}(y_{2}|y)p_{Y|X_{1},X_{2}}(y|x_{1},x_{2})p_{X_{1}}(x_{1})p_{X_{2}}(x_{2}).
\end{equation}

Then, the mutual information term $I(X_{1};Y|X_{2})$ is
\begin{equation}\label{appeee5}
I(X_{1};Y|X_{2})=(1-\beta)h(\alpha)\leq h(\alpha)\leq 1,
\end{equation}
where $h(\alpha)=-\alpha\log(\alpha)-(1-\alpha)\log(1-\alpha)$.
Similarly, $I(X_{2};Y|X_{1})\leq h(\beta)\leq 1$, $I(X_{1},X_{2};Y)=h(\alpha+\beta-\alpha \beta)\leq 1$,
$I(X_{2};Y|X_{1})-I(X_{2};Y_{1}|X_{1})\leq h(\beta)+h(p)-h(\beta+p-2\beta p)\leq h(p)$
and $I(X_{1};Y|X_{2})-I(X_{1};Y_{2}|X_{2})\leq h(\alpha)+h(q)-h(\alpha+q-2\alpha p)\leq h(q)$.

Theorem \ref{T3} and Theorem \ref{T3x} are obtained.

\renewcommand{\theequation}{\arabic{equation}}
\section{Proof of the Converse Part of Theorem \ref{T4}\label{appen5}}

In this section, we establish the converse part of Theorem \ref{T4}: all the achievable $(R_{1}, R_{2}, R_{e})$ triples are
contained in the set $\mathcal{R^{(D)}}$. We will prove the inequalities in Theorem \ref{T4} in the remaining of this section.

\textbf{(Proof of $0\leq R_{1}\leq I(V;Y|U_{2})$)} The proof of this inequality is as follows.
\begin{eqnarray}\label{appey1}
\frac{1}{N}H(W_{1})&\stackrel{(a)}\leq&\frac{1}{N}(H(W_{1}|W_{2})+\delta(P_{e})-H(W_{1}|W_{2},Y^{N}))\nonumber\\
&=&\frac{1}{N}(I(W_{1};Y^{N}|W_{2})+\delta(P_{e}))\nonumber\\
&=&\frac{1}{N}\sum_{i=1}^{N}(H(Y_{i}|Y^{i-1},W_{2})-H(Y_{i}|Y^{i-1},W_{2},W_{1}))+\frac{\delta(P_{e})}{N}\nonumber\\
&\leq&\frac{1}{N}\sum_{i=1}^{N}(H(Y_{i}|Y^{i-1},W_{2})-H(Y_{i}|Y^{i-1},W_{2},W_{1},Z_{i+1}^{N}))+\frac{\delta(P_{e})}{N}\nonumber\\
&\stackrel{(b)}=&\frac{1}{N}\sum_{i=1}^{N}(H(Y_{i}|U_{2,i})-H(Y_{i}|U_{2,i},V_{i}))+\frac{\delta(P_{e})}{N}\nonumber\\
&\stackrel{(c)}=&I(V;Y|U_{2})+\frac{\delta(P_{e})}{N},
\end{eqnarray}
where (a) is from the Fano's inequality and the fact that $W_{1}$ is independent of $W_{2}$, (b) is from
the definitions $U_{2,i}\triangleq (Y^{i-1},W_{2})$ and $V_{i}\triangleq (Y^{i-1},W_{2},W_{1},Z_{i+1}^{N})$,
and (c) is from the definitions that $U_{2}\triangleq (U_{2,J}, J)$, $V\triangleq (V_{J}, J)$, $Y\triangleq Y_{J}$,
where $J$ is a random variable (uniformly distributed over $\{1,2,...,N\}$), and it is independent of $U_{2,i}$, $V_{i}$ and $Y_{i}$.

By using $P_{e}\leq \epsilon$, $R_{1}=\lim_{N\rightarrow \infty}\frac{H(W_{1})}{N}$ and (\ref{appey1}), it is easy to see that
$0\leq R_{1}\leq I(V;Y|U_{2})$.

\textbf{(Proof of $0\leq R_{2}\leq I(V;Y|U_{1})$)} The proof is similar to the proof of $0\leq R_{1}\leq I(V;Y|U_{2})$, and it is omitted here.
Note that $U_{1}\triangleq (Y^{J-1},W_{1}, J)$.

\textbf{(Proof of $0\leq R_{1}+R_{2}\leq I(V;Y)$)}
\begin{eqnarray}\label{appey2}
\frac{1}{N}H(W_{1},W_{2})&\stackrel{(1)}\leq&\frac{1}{N}(I(W_{1},W_{2};Y^{N})+\delta(P_{e}))\nonumber\\
&=&\frac{1}{N}\sum_{i=1}^{N}(H(Y_{i}|Y^{i-1})-H(Y_{i}|Y^{i-1},W_{1},W_{2}))+\frac{\delta(P_{e})}{N}\nonumber\\
&\leq&\frac{1}{N}\sum_{i=1}^{N}(H(Y_{i})-H(Y_{i}|Y^{i-1},W_{1},W_{2},Z_{i+1}^{N}))+\frac{\delta(P_{e})}{N}\nonumber\\
&\stackrel{(2)}=&\frac{1}{N}\sum_{i=1}^{N}(H(Y_{i})-H(Y_{i}|V_{i}))+\frac{\delta(P_{e})}{N}\nonumber\\
&\stackrel{(3)}=&I(V;Y)+\frac{\delta(P_{e})}{N},
\end{eqnarray}
where (1) is from the Fano's inequality, (2) is from the definition $V_{i}\triangleq (Y^{i-1},W_{2},W_{1},Z_{i+1}^{N})$,
and (3) is from the definitions that $V\triangleq (V_{J}, J)$, $Y\triangleq Y_{J}$,
where $J$ is a random variable (uniformly distributed over $\{1,2,...,N\}$), and it is independent of $V_{i}$ and $Y_{i}$.

By using $P_{e}\leq \epsilon$, $R_{1}=\lim_{N\rightarrow \infty}\frac{H(W_{1})}{N}$, $R_{2}=\lim_{N\rightarrow \infty}\frac{H(W_{2})}{N}$
and (\ref{appey2}), it is easy to see that
$0\leq R_{1}+R_{2}\leq I(V;Y)$.

\textbf{(Proof of $0\leq R_{e}\leq R_{1}+R_{2}$)} This inequality is obtained by the following (\ref{appey3}).
\begin{equation}\label{appey3}
R_{e}\leq\lim_{N\rightarrow \infty}\frac{1}{N}H(W_{1},W_{2}|Y^{N})\leq \lim_{N\rightarrow \infty}\frac{1}{N}H(W_{1},W_{2})=R_{1}+R_{2}.
\end{equation}

\textbf{(Proof of $R_{e}\leq I(V;Y|U)-I(V;Z|U)$)} The proof is obtained by substituting (\ref{appey5}), (\ref{appey6}),
(\ref{appey7}) and (\ref{appey10}) into (\ref{appey4}),
and using $P_{e}\leq \epsilon$ and the definitions $U\triangleq (Y^{J-1};Z_{J+1}^{N},J)$ and $V\triangleq (W_{1},W_{2},Y^{J-1};Z_{J+1}^{N},J)$.

\begin{eqnarray}\label{appey4}
R_{e}&\leq&\lim_{N\rightarrow \infty}\frac{1}{N}H(W_{1},W_{2}|Z^{N})\nonumber\\
&=&\lim_{N\rightarrow \infty}\frac{1}{N}(H(W_{1},W_{2})-I(W_{1},W_{2};Z^{N}))\nonumber\\
&=&\lim_{N\rightarrow \infty}\frac{1}{N}(I(W_{1},W_{2};Y^{N})+H(W_{1},W_{2}|Y^{N})-I(W_{1},W_{2};Z^{N}))\nonumber\\
&\stackrel{(a)}\leq&\lim_{N\rightarrow \infty}\frac{1}{N}(I(W_{1},W_{2};Y^{N})-I(W_{1},W_{2};Z^{N})+\delta(P_{e})),
\end{eqnarray}
where (a) is from the Fano's inequality.

\begin{eqnarray}\label{appey5}
I(W_{1},W_{2};Y^{N})&=&\sum_{i=1}^{N}(H(Y_{i}|Y^{i-1})-H(Y_{i}|Y^{i-1},W_{1},W_{2}))\nonumber\\
&=&\sum_{i=1}^{N}(H(Y_{i}|Y^{i-1})-H(Y_{i}|Y^{i-1},W_{1},W_{2})-H(Y_{i}|Y^{i-1},Z_{i+1}^{N})+H(Y_{i}|Y^{i-1},Z_{i+1}^{N})\nonumber\\
&&-H(Y_{i}|Y^{i-1},W_{1},W_{2},Z_{i+1}^{N})+H(Y_{i}|Y^{i-1},W_{1},W_{2},Z_{i+1}^{N}))\nonumber\\
&=&\sum_{i=1}^{N}(I(Y_{i};Z_{i+1}^{N}|Y^{i-1})-I(Y_{i};Z_{i+1}^{N}|Y^{i-1},W_{1},W_{2})+I(Y_{i};W_{1},W_{2}|Y^{i-1},Z_{i+1}^{N})).
\end{eqnarray}

\begin{eqnarray}\label{appey6}
I(W_{1},W_{2};Z^{N})&=&\sum_{i=1}^{N}(H(Z_{i}|Z_{i+1}^{N})-H(Z_{i}|Z_{i+1}^{N},W_{1},W_{2}))\nonumber\\
&=&\sum_{i=1}^{N}(H(Z_{i}|Z_{i+1}^{N})-H(Z_{i}|Z_{i+1}^{N},W_{1},W_{2})-H(Z_{i}|Y^{i-1},Z_{i+1}^{N})+H(Z_{i}|Y^{i-1},Z_{i+1}^{N})\nonumber\\
&&-H(Z_{i}|Y^{i-1},W_{1},W_{2},Z_{i+1}^{N})+H(Z_{i}|Y^{i-1},W_{1},W_{2},Z_{i+1}^{N}))\nonumber\\
&=&\sum_{i=1}^{N}(I(Z_{i};Y^{i-1}|Z_{i+1}^{N})-I(Z_{i};Y^{i-1}|Z_{i+1}^{N},W_{1},W_{2})+I(Z_{i};W_{1},W_{2}|Y^{i-1},Z_{i+1}^{N})).
\end{eqnarray}

Note that
\begin{equation}\label{appey7}
\sum_{i=1}^{N}I(Z_{i};Y^{i-1}|Z_{i+1}^{N})=\sum_{i=1}^{N}I(Y_{i};Z_{i+1}^{N}|Y^{i-1}).
\end{equation}

\begin{IEEEproof}[Proof of (\ref{appey7})]
The right hand side of (\ref{appey7}) is equal to
\begin{equation}\label{appey8}
\sum_{i=1}^{N}I(Y_{i};Z_{i+1}^{N}|Y^{i-1})=\sum_{i=1}^{N}\sum_{j=i+1}^{N}I(Y_{i};Z_{j}|Y^{i-1},Z_{j+1}^{N}),
\end{equation}
and the left hand side of (\ref{appey7}) is equal to
\begin{eqnarray}\label{appey9}
\sum_{i=1}^{N}I(Z_{i};Y^{i-1}|Z_{i+1}^{N})&=&\sum_{i=1}^{N}\sum_{j=1}^{i-1}I(Z_{i};Y_{j}|Z_{i+1}^{N},Y^{j-1})\nonumber\\
&=&\sum_{j=1}^{N}\sum_{i=1}^{j-1}I(Z_{j};Y_{i}|Z_{j+1}^{N},Y^{i-1})\nonumber\\
&=&\sum_{i=1}^{N}\sum_{j=i+1}^{N}I(Z_{j};Y_{i}|Z_{j+1}^{N},Y^{i-1}),
\end{eqnarray}
and therefore, the formula (\ref{appey7}) is verified by (\ref{appey8}) and (\ref{appey9}).
\end{IEEEproof}

Analogously,
\begin{equation}\label{appey10}
\sum_{i=1}^{N}I(Y_{i};Z_{i+1}^{N}|Y^{i-1},W_{1},W_{2})=\sum_{i=1}^{N}I(Z_{i};Y^{i-1}|Z_{i+1}^{N},W_{1},W_{2}).
\end{equation}

The Markov chain $(U,U_{1},U_{2})\rightarrow V\rightarrow (X_{1},X_{2})\rightarrow (Y,Z)$ is directly obtained from the above definitions.

The proof of the converse part of Theorem \ref{T4} is completed.

\renewcommand{\theequation}{\arabic{equation}}
\section{Proof of the Direct Part of Theorem \ref{T4}\label{appen5a}}

In this section we establish the direct part of
Theorem \ref{T4}(about existence). Suppose
$(R_{1},R_{2},R_{e})\in \mathcal{R^{D}}$, we will show that $(R_{1},R_{2},R_{e})$ is achievable.

The existence of the encoder-decoder is under the sufficient condition $R_{e}=I(V;Y|U)-I(V;Z|U)$.
Given a triple $(R_{1},R_{2},R_{e})$, choose a joint probability mass function $p_{U,U_{1},U_{2},V,X_{1},X_{2},Y,Z}(u,u_{1},u_{2},v,x_{1},x_{2},y,z)$
such that
$$0\leq R_{1}\leq I(V;Y|U_{2}), 0\leq R_{2}\leq I(V;Y|U_{1}), R_{1}+R_{2}\leq I(V;Y),$$
$$R_{e}\leq R_{1}+R_{2}, \ \ R_{e}=I(V;Y|U)-I(V;Z|U).$$

The message sets $\mathcal{W}_{1}$ and $\mathcal{W}_{2}$ satisfy the following conditions:
\begin{equation}\label{ao1}
\lim_{N\rightarrow \infty}\frac{1}{N}\log\parallel \mathcal{W}_{1}\parallel=R_{1},
\end{equation}
\begin{equation}\label{ao2}
\lim_{N\rightarrow \infty}\frac{1}{N}\log\parallel \mathcal{W}_{2}\parallel= R_{2}.
\end{equation}
Note that
\begin{equation}\label{ao3}
\lim_{N\rightarrow \infty}\frac{1}{N}\log\parallel \mathcal{W}_{1}\times \mathcal{W}_{2}\parallel= R_{1}+R_{2}\geq I(V;Y|U)-I(V;Z|U).
\end{equation}

Now the
remaining of this section is organized as follows.
The encoding-decoding scheme is
introduced in Subsection \ref{app6.2}. For any given $\epsilon> 0$, the
proofs of $\lim_{N\rightarrow \infty}\frac{\log\parallel \mathcal{W}_{1}\parallel}{N}= R_{1}$,
$\lim_{N\rightarrow \infty}\frac{\log\parallel \mathcal{W}_{2}\parallel}{N}= R_{2}$,
$\lim_{N\rightarrow \infty}\Delta\geq R_{e}$
and $P_{e}\leq \epsilon$ are given in Subsection \ref{app6.3}.

\begin{figure}[htb]
\centering
\includegraphics[scale=0.6]{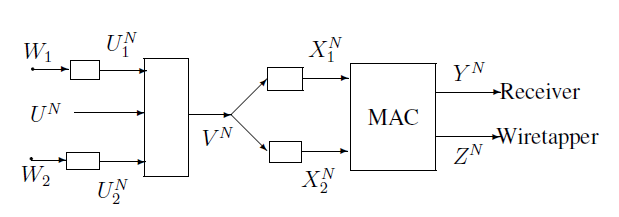}
\caption{Code construction for MAC-WT with cooperative encoders}
\label{fx}
\end{figure}

\subsection{Encoding-decoding Scheme\label{app6.2}}

The encoding scheme for the MAC-WT with cooperative encoders is in Figure \ref{fx}. In the reminder of this subsection,
we will introduce the realization of the random vectors in Figure \ref{fx}.

\begin{itemize}

\item (\textbf{A realization of $U_{1}^{N}$ and $U_{2}^{N}$})
For each $w_{1}$ ($w_{1}\in \{1,2,...,2^{NR_{1}}\}$), generate a corresponding codeword $u_{1}^{N}(w_{1})$ i.i.d. according
to the probability mass function $p_{U_{1}}(u_{1})$. Similarly,
for each $w_{2}$ ($w_{1}\in \{1,2,...,2^{NR_{2}}\}$), generate a corresponding codeword $u_{2}^{N}(w_{2})$ i.i.d. according
to the probability mass function $p_{U_{2}}(u_{2})$.
$u_{1}^{N}(w_{1})$ and $u_{2}^{N}(w_{2})$ are realizations of the random vectors $U_{1}^{N}$ and $U_{2}^{N}$, respectively.

\item (\textbf{A realization of $U^{N}$})
Let $u^{N}(m)$ ($1\leq m\leq 2^{N\gamma}$) be chosen from the strong typical set $T^{N}_{U}(\eta^{*})$,
where $\eta^{*}$ is an arbitrary small positive real number and $0\leq \gamma\leq \min\{I(U;Y),I(U;Z)\}$.
Moreover, let $\Phi$ be a set defined as
$\Phi=\{u^{N}(m): 1\leq m\leq 2^{N\gamma}\}$. Note that the elements of $\Phi$
are distinguishable.
Choose a sequence $u^{N}(m)$ from the set $\Phi$ as a realization of $U^{N}$, and label the sequence $u^{N}(m)$
as $m$.

\item (Step ii) (\textbf{A realization of $V^{N}$})

Let $J$, $L$ and $M$ be the random variables used for indexing the random vector $V^{N}$,
and the three random variables take values in the index sets $\mathcal{J}$, $\mathcal{L}$ and $\mathcal{M}$, respectively.
Let $v_{jlm}^{N}$ be a realization of the random vector $V^{N}$,
where $j$, $l$, $m$ run over the index sets
$\mathcal{J}$, $\mathcal{L}$ and $\mathcal{M}$.
The construction of the sequence $v_{jlm}^{N}$ is considered in three parts. The first part is about the
determination of the size of the indices
$j$, $l$ and $m$ appeared in the sequence $v_{jlm}^{N}$.
The second part is the full details of how to choose the indices of the sequence $v_{jlm}^{N}$.
The third part is the construction of the sequence $v_{jlm}^{N}$, see the following.

\begin{itemize}

\item (\textbf{The size of $\mathcal{J}$, $\mathcal{L}$ and $\mathcal{M}$})

The indices
$j$, $l$ and $m$ appeared in the sequence $v_{jlm}^{N}$ respectively run over the index sets
$\mathcal{J}$, $\mathcal{L}$, $\mathcal{M}$ with the following properties:
\begin{equation}\label{ao4}
\lim_{N\rightarrow \infty}\frac{1}{N}\log\parallel \mathcal{J}\parallel=I(V;Z|U),
\end{equation}
\begin{equation}\label{ao5}
\lim_{N\rightarrow \infty}\frac{1}{N}\log\parallel \mathcal{L}\parallel=I(V;Y|U)-I(V;Z|U),
\end{equation}
\begin{equation}\label{ao6}
\lim_{N\rightarrow \infty}\frac{1}{N}\log\parallel \mathcal{M}\parallel=\gamma,
\end{equation}
where $\gamma$ satisfies $0\leq \gamma\leq \min\{I(U;Y),I(U;Z)\}$.

\item (\textbf{The chosen of $j$, $l$ and $m$})

\begin{itemize}

\item (Case 1) If $R_{1}+R_{2}\geq I(V;Y|U)$, let $\mathcal{W}_{1}\times \mathcal{W}_{2}=\mathcal{J}\times \mathcal{L}\times \mathcal{M}$.
Therefore, in this case, the chosen of $u^{N}(m)$ is based on $w_{1}$ and $w_{2}$.

The indices $j$, $l$ and $m$ are chosen based on $w_{1}$ and $w_{2}$.

\item (Case 2) If $R_{1}+R_{2}\leq I(V;Y|U)$, let $\mathcal{W}_{1}\times \mathcal{W}_{2}=\mathcal{L}\times \mathcal{K}$, where $\mathcal{K}$
is an arbitrary set such that (\ref{ao3}) holds.
Let $\bar{g}$ be a mapping of $\mathcal{J}$ into $\mathcal{K}$, partitioning  $\mathcal{J}$ into subsets of nearly equal size.
Note that in this case, the chosen of $u^{N}(m)$ is not based on $w_{1}$ and $w_{2}$.

The index $j$ is randomly chosen from the set
$\bar{g}^{-1}(k)\subset \mathcal{J}$ (where $\bar{g}^{-1}$ is the inverse mapping of $\bar{g}$, and $k\in \mathcal{K}$).

The index $m$ is chosen according to the label of $u^{N}(m)$.

The index $l$ is chosen from $\mathcal{L}$.

\end{itemize}

\item (\textbf{The construction of $v_{jlm}^{N}$}) The construction of $v_{jlm}^{N}$ is as follows.
For each $m\in \mathcal{M}$, there exists a $U$-typical sequence $u^{N}(m)\in \Phi$ such that
all the $v_{jlm}^{N}$ are $V|U$-generated by $u^{N}(m)$,
and this indicates that $v_{jlm}^{N}\in T^{N}_{V}(\eta^{**})$,
where $\eta^{**}$ is an arbitrary small positive real number.
\end{itemize}

\item (\textbf{A realization of $X_{1}^{N}$ and $X_{2}^{N}$}) $x_{1}^{N}$ is generated
according to a new discrete memoryless channel (DMC) with input $v_{jlm}^{N}$
and output $x_{1}^{N}$. The transition probability of this new DMC is $p_{X_{1}|V}(x_{1}|v)$.

Similarly, $x_{2}^{N}$ is generated
according to a new discrete memoryless channel (DMC) with input $v_{jlm}^{N}$
and output $x_{2}^{N}$. The transition probability of this new DMC is $p_{X_{2}|V}(x_{2}|v)$.

\item (\textbf{Decoding scheme of the legitimate receiver})
For given $y^{N}$, try to find a sequence $v_{\hat{j}\hat{l}\hat{m}}^{N}$ such that
$(v_{\hat{j}\hat{l}\hat{m}}^{N},y^{N})\in T^{N}_{VY}(\epsilon^{***})$.
If there exist sequences with the same indices $\hat{j}$, $\hat{l}$ and $\hat{m}$, put out the corresponding
$\hat{w}_{1}$ and $\hat{w}_{2}$, else declare a decoding error.

\end{itemize}

\subsection{Achievability Proof\label{app6.3}}

By using the above definitions, it is easy to verify that  $\lim_{N\rightarrow \infty}\frac{\log\parallel \mathcal{W}_{1}\parallel}{N}= R_{1}$ and
$\lim_{N\rightarrow \infty}\frac{\log\parallel \mathcal{W}_{2}\parallel}{N}= R_{2}$.

Then, observing the construction of
$V^{N}$, it is easy to see that the codewords of $V^{N}$ are upper-bounded
by $2^{NI(V;Y)}$. Therefore, from the standard channel coding theorem, for any given $\epsilon>0$ and sufficiently large $N$, we have
$P_{e}\leq \epsilon$.

It remains to show that $\lim_{N\rightarrow \infty}\Delta\geq R_{e}$, see the following. Let $M$ be the random variable
defined as the third coordinate of the actual value of $V^{N}$. Then
\begin{eqnarray}\label{ao67}
\lim_{N\rightarrow \infty}\Delta&\triangleq&\lim_{N\rightarrow \infty}\frac{1}{N}H(W_{1},W_{2}|Z^{N})\nonumber\\
&\geq&\lim_{N\rightarrow \infty}\frac{1}{N}H(W_{1},W_{2}|Z^{N},M)\nonumber\\
&=&\lim_{N\rightarrow \infty}\frac{1}{N}(H(W_{1},W_{2},Z^{N}|M)-H(Z^{N}|M))\nonumber\\
&=&\lim_{N\rightarrow \infty}\frac{1}{N}(H(W_{1},W_{2},V^{N},Z^{N}|M)-H(V^{N}|W_{1},W_{2},Z^{N},M)-H(Z^{N}|M))\nonumber\\
&\stackrel{(a)}=&\lim_{N\rightarrow \infty}\frac{1}{N}(H(W_{1},W_{2},V^{N}|M)+H(Z^{N}|V^{N})-H(V^{N}|W_{1},W_{2},Z^{N},M)-H(Z^{N}|M))\nonumber\\
&\geq&\lim_{N\rightarrow \infty}\frac{1}{N}(H(V^{N}|M)+H(Z^{N}|V^{N})-H(V^{N}|W_{1},W_{2},Z^{N},M)-H(Z^{N}|M)),
\end{eqnarray}
where (a) is from the Markov chain $(W_{1},W_{2},M)\rightarrow V^{N}\rightarrow Z^{N}$.

The first term in (\ref{ao67}) can be bounded as follows.
\begin{equation}\label{ao68}
\lim_{N\rightarrow \infty}\frac{1}{N}H(V^{N}|M)\geq I(V;Y|U),
\end{equation}
where (\ref{ao68}) is from the property of the strong typical sequences and the construction of $V^{N}$, see \cite[p. 343]{CK}.

The second term in (\ref{ao67}) is as follows.
\begin{equation}\label{ao69}
\lim_{N\rightarrow \infty}\frac{1}{N}H(Z^{N}|V^{N})\geq H(Z|V),
\end{equation}
where (\ref{ao69}) is from a similar proof in \cite[p. 343]{CK}.

For the third term in (\ref{ao67}), we have
\begin{equation}\label{ao70}
\lim_{N\rightarrow \infty}\frac{1}{N}H(V^{N}|W_{1},W_{2},Z^{N},M)=0.
\end{equation}
This is because for given $m$, $w_{1}$ and $w_{2}$, there are at most $2^{NI(V;Z|U)}$ codewords left for $v^{N}$.
Then note that
\begin{eqnarray*}
I(V;Z|U)&=&H(Z|U)-H(Z|V)\nonumber\\
&\leq&I(V;Z).
\end{eqnarray*}
From the standard channel coding theorem and the Fano's inequality,
we have (\ref{ao70}).

For the fourth term in (\ref{ao67}), we have
\begin{equation}\label{ao71}
\lim_{N\rightarrow \infty}\frac{1}{N}H(Z^{N}|M)\leq H(Z|U),
\end{equation}
and this is from a standard technique as in \cite[p. 343]{CK}.

Substituting (\ref{ao68}), (\ref{ao69}), (\ref{ao70}) and (\ref{ao71}) into (\ref{ao67}), we have
\begin{equation}\label{ao72}
\lim_{N\rightarrow \infty}\Delta\geq I(V;Y|U)-I(V;Z|U).
\end{equation}

Therefore, the achievability proof for Theorem \ref{T4} is completed.

\renewcommand{\theequation}{\arabic{equation}}
\section{Size Constraints of the Auxiliary Random Variables in Theorem \ref{T4}\label{appen7}}

By using the support lemma (see \cite{Cs}, p.310), it suffices to show that the random variables
$U_{1}$, $U_{2}$, $U$ and $V$ can be replaced by new ones, preserving the Markovity
$(U,U_{1},U_{2})\rightarrow V\rightarrow (X_{1}, X_{2})\rightarrow (Y,Z)$ and the
mutual information $I(V;Y|U_{1})$, $I(V;Y|U_{2})$, $I(V;Y)$, $I(V;Y|U)$,
$I(V;Z|U)$, and furthermore,
the ranges of the new $U_{1}$, $U_{2}$, $U$ and $V$ satisfy:
$$\|\mathcal{U}\|\leq \|\mathcal{X}_{1}\|\|\mathcal{X}_{2}\|+1,$$
$$\|\mathcal{U}_{1}\|\leq \|\mathcal{X}_{1}\|\|\mathcal{X}_{2}\|,$$
$$\|\mathcal{U}_{2}\|\leq \|\mathcal{X}_{1}\|\|\mathcal{X}_{2}\|,$$
$$\|\mathcal{V}\|\leq (\|\mathcal{X}_{1}\|\|\mathcal{X}_{2}\|+1)^{2}\|\mathcal{X}_{1}\|^{2}\|\mathcal{X}_{2}\|^{2}.$$
The proof is in the reminder of this section.

Let
\begin{equation}\label{e601.1}
\bar{p}=p_{X_{1}X_{2}}(x_{1},x_{2}).
\end{equation}
Define the following continuous scalar functions of $\bar{p}$ :
$$f_{X_{1}X_{2}}(\bar{p})=p_{X_{1}X_{2}}(x_{1},x_{2}), \ \ f_{Y}(\bar{p})=H(Y), \ \ f_{Z}(\bar{p})=H(Z).$$
Since there are $\|\mathcal{X}_{1}\|\|\mathcal{X}_{2}\|-1$ functions of
$f_{X_{1}X_{2}}(\bar{p})$, the total number of the continuous scalar
functions of $\bar{p}$ is $\|\mathcal{X}_{1}\|\|\mathcal{X}_{2}\|$+1.

Let $\bar{p}_{X_{1}X_{2}|U}=Pr\{X_{1}=x_{1},X_{2}=x_{2}|U=u\}$. With these distributions
$\bar{p}_{X_{1}X_{2}|U}$, we have
\begin{equation}\label{e602.1}
p_{X_{1}X_{2}}(x_{1},x_{2})=\sum_{u\in \mathcal{U}}p(U=u)f_{X_{1}X_{2}}(\bar{p}_{X_{1}X_{2}|U}),
\end{equation}
\begin{equation}\label{e603.1}
H(Y|U)=\sum_{u\in \mathcal{U}}p(U=u)f_{Y}(\bar{p}_{X_{1}X_{2}|U}),
\end{equation}
\begin{equation}\label{e605.1}
H(Z|U)=\sum_{u\in
\mathcal{U}}p(U=u)f_{Z}(\bar{p}_{X_{1}X_{2}|U}).
\end{equation}

According to the support lemma (\cite{Cs}, p.310), the random
variable $U$ can be replaced by new ones such that the new $U$ takes
at most $\|\mathcal{X}_{1}\|\|\mathcal{X}_{2}\|+1$ different values and the
expressions (\ref{e602.1}), (\ref{e603.1}) and (\ref{e605.1}) are preserved.

Similarly, we can prove that $\|\mathcal{U}_{1}\|\leq \|\mathcal{X}_{1}\|\|\mathcal{X}_{2}\|$ and
$\|\mathcal{U}_{2}\|\leq \|\mathcal{X}_{1}\|\|\mathcal{X}_{2}\|$.

Once the alphabets of $U$, $U_{1}$, $U_{2}$ are fixed, we apply similar arguments to bound the alphabet of $V$, see the following.
Define $\|\mathcal{X}_{1}\|\|\mathcal{X}_{2}\|+1$ continuous scalar functions of $\bar{p}_{X_{1}X_{2}}$ :
$$f_{X_{1}X_{2}}(\bar{p}_{X_{1}X_{2}})=p_{X_{1}X_{2}}(x_{1},x_{2}), f_{Y}(\bar{p}_{X_{1}X_{2}})=H(Y), f_{Z}(\bar{p}_{X_{1}X_{2}})=H(Z),$$
where of the functions $f_{X_{1}X_{2}}(\bar{p}_{X_{1}X_{2}})$, only $\|\mathcal{X}_{1}\|\|\mathcal{X}_{2}\|-1$ are to be considered.

For fixed $u$, $u_{1}$ and $u_{2}$, let $\bar{p}_{X_{1}X_{2}|V}=Pr\{X_{1}=x_{1},X_{2}=x_{2}|V=v\}$. With these distributions $\bar{p}_{X_{1}X_{2}|V}$, we have
\begin{equation}\label{e607.1}
Pr\{X_{1}=x_{1},X_{2}=x_{2}|U=u,U_{1}=u_{1},U_{2}=u_{2}\}=\sum_{v\in \mathcal{V}}Pr\{V=v|U=u,U_{1}=u_{1},U_{2}=u_{2}\}f_{X_{1}X_{2}}(\bar{p}_{X_{1}X_{2}|V}),
\end{equation}
\begin{equation}\label{e608.1}
H(Y|V)=\sum_{v\in \mathcal{V}}f_{Y}(\bar{p}_{X_{1}X_{2}|V})Pr\{V=v|U=u,U_{1}=u_{1},U_{2}=u_{2}\},
\end{equation}
\begin{equation}\label{e609.1}
H(Z|V)=\sum_{v\in \mathcal{V}}f_{Z}(\bar{p}_{X_{1}X_{2}|V})Pr\{V=v|U=u,U_{1}=u_{1},U_{2}=u_{2}\}.
\end{equation}

By the support lemma (\cite{Cs}, p.310), for fixed $u$, $u_{1}$ and $u_{2}$, the size of the alphabet of the random variable $V$ can not be larger than
$\|\mathcal{X}_{1}\|\|\mathcal{X}_{2}\|+1$, and therefore,
$\|\mathcal{V}\|\leq (\|\mathcal{X}_{1}\|\|\mathcal{X}_{2}\|+1)^{2}\|\mathcal{X}_{1}\|^{2}\|\mathcal{X}_{2}\|^{2}$ is proved.

\renewcommand{\theequation}{\arabic{equation}}
\section{Proof of Theorem \ref{T5}\label{appen8}}

The only difference between Theorem \ref{T4} and Theorem \ref{T5} is the upper bound of $R_{e}$.
Since the degraded MAC-WT with cooperative encoders is a special case of the general model, and therefore,
the converse proof of Theorem \ref{T5} can be directly obtained from the converse
proof of Theorem \ref{T4} and (\ref{e50aa}). Now it remains to prove the achievability, see the remainder of this section.

\begin{figure}[htb]
\centering
\includegraphics[scale=0.6]{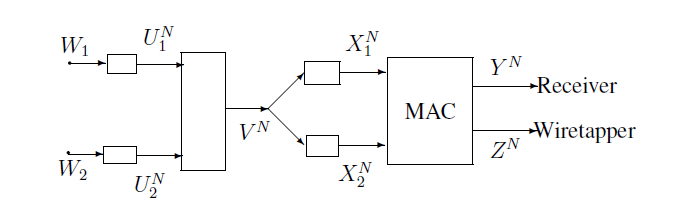}
\caption{Code construction for degraded MAC-WT with cooperative encoders}
\label{fxx}
\end{figure}

The encoding-decoding scheme for Theorem \ref{T5} is a special case of that for Theorem \ref{T4},
see Figure \ref{fxx}.

The existence of the encoder-decoder is under the sufficient condition $R_{e}=I(V;Y)-I(V;Z)$.
Given a triple $(R_{1},R_{2},R_{e})$, choose a joint probability mass function $p_{U_{1},U_{2},V,X_{1},X_{2},Y,Z}(u_{1},u_{2},v,x_{1},x_{2},y,z)$
such that
$$0\leq R_{1}\leq I(V;Y|U_{2}), 0\leq R_{2}\leq I(V;Y|U_{1}), R_{1}+R_{2}\leq I(V;Y),$$
$$R_{e}\leq R_{1}+R_{2}, R_{e}=I(V;Y)-I(V;Z).$$

The message sets $\mathcal{W}_{1}$ and $\mathcal{W}_{2}$ satisfy the following conditions:
\begin{equation}\label{aox1}
\lim_{N\rightarrow \infty}\frac{1}{N}\log\parallel \mathcal{W}_{1}\parallel=R_{1},
\end{equation}
\begin{equation}\label{aox2}
\lim_{N\rightarrow \infty}\frac{1}{N}\log\parallel \mathcal{W}_{2}\parallel= R_{2}.
\end{equation}
Note that
\begin{equation}\label{aox3}
\lim_{N\rightarrow \infty}\frac{1}{N}\log\parallel \mathcal{W}_{1}\times \mathcal{W}_{2}\parallel= R_{1}+R_{2}\geq I(V;Y)-I(V;Z).
\end{equation}

The construction of $U_{1}^{N}$ and $U_{2}^{N}$ in Figure \ref{fxx} is the same as those in Appendix \ref{appen7}, and
$V^{N}$ is constructed as follows.

Generate $2^{N(I(V;Y)-\epsilon_{N})}$ codewords $v^{N}$ ($\epsilon_{N}\rightarrow \infty$ as $N\rightarrow \infty$),
and each of them is uniformly drawn from the strong typical set
$T^{N}_{V}(\eta)$. Divide the $2^{N(I(V;Y)-\epsilon_{N})}$ codewords into $2^{N(R_{1}+R_{2})}$ bins, and each bin corresponds to
a specific value in $\mathcal{W}_{1}\times \mathcal{W}_{2}$.

(\textbf{A realization of $X_{1}^{N}$ and $X_{2}^{N}$}) $x_{1}^{N}$ is generated
according to a new discrete memoryless channel (DMC) with input $v_{jlm}^{N}$
and output $x_{1}^{N}$. The transition probability of this new DMC is $p_{X_{1}|V}(x_{1}|v)$.

Similarly, $x_{2}^{N}$ is generated
according to a new discrete memoryless channel (DMC) with input $v_{jlm}^{N}$
and output $x_{2}^{N}$. The transition probability of this new DMC is $p_{X_{2}|V}(x_{2}|v)$.

(\textbf{Decoding scheme of the legitimate receiver})
For given $y^{N}$, try to find a sequence $v^{N}(\hat{w}_{1},\hat{w}_{2})$ such that
$(v^{N}(\hat{w}_{1},\hat{w}_{2}),y^{N})\in T^{N}_{VY}(\epsilon^{***})$.
If there exist sequences with the same $\hat{w}_{1}$ and $\hat{w}_{2}$, put out the corresponding
$\hat{w}_{1}$ and $\hat{w}_{2}$, else declare a decoding error.

By using the above definitions, it is easy to verify that  $\lim_{N\rightarrow \infty}\frac{\log\parallel \mathcal{W}_{1}\parallel}{N}= R_{1}$ and
$\lim_{N\rightarrow \infty}\frac{\log\parallel \mathcal{W}_{2}\parallel}{N}= R_{2}$.

Then, observing the construction of
$V^{N}$, it is easy to see that the codewords of $V^{N}$ are upper-bounded
by $2^{NI(V;Y)}$. Therefore, from the standard channel coding theorem, for any given $\epsilon>0$ and sufficiently large $N$, we have
$P_{e}\leq \epsilon$.

It remains to show that $\lim_{N\rightarrow \infty}\Delta\geq R_{e}$, see the following.
\begin{eqnarray}\label{aox4}
\lim_{N\rightarrow \infty}\Delta&\triangleq&\lim_{N\rightarrow \infty}\frac{1}{N}H(W_{1},W_{2}|Z^{N})\nonumber\\
&=&\lim_{N\rightarrow \infty}\frac{1}{N}(H(W_{1},W_{2},Z^{N})-H(Z^{N}))\nonumber\\
&=&\lim_{N\rightarrow \infty}\frac{1}{N}(H(W_{1},W_{2},V^{N},Z^{N})-H(V^{N}|W_{1},W_{2},Z^{N})-H(Z^{N}))\nonumber\\
&=&\lim_{N\rightarrow \infty}\frac{1}{N}(H(V^{N}|W_{1},W_{2})+H(W_{1},W_{2})-H(V^{N}|W_{1},W_{2},Z^{N})-I(V^{N};Z^{N})).
\end{eqnarray}

The first term in (\ref{aox4}) is
\begin{equation}\label{aox5}
\lim_{N\rightarrow \infty}\frac{1}{N}H(V^{N}|W_{1},W_{2})=I(V;Y)-R_{1}-R_{2}.
\end{equation}

The second term in (\ref{aox4}) is as follows.
\begin{equation}\label{aox6}
\lim_{N\rightarrow \infty}\frac{1}{N}H(W_{1},W_{2})=R_{1}+R_{2}.
\end{equation}

For the third term in (\ref{aox4}), we have
\begin{equation}\label{aox7}
\lim_{N\rightarrow \infty}\frac{1}{N}H(V^{N}|W_{1},W_{2},Z^{N})=0.
\end{equation}
This is because for given $w_{1}$ and $w_{2}$, there are $2^{N(I(V;Y)-\epsilon_{N}-R_{1}-R_{2})}$ codewords left for $v^{N}$.
Then note that
\begin{eqnarray*}
I(V;Y)-\epsilon_{N}-R_{2}-R_{1}&\leq&I(V;Y)-\epsilon_{N}-R_{e}\nonumber\\
&=&I(V;Y)-\epsilon_{N}-(I(V;Y)-I(V;Z))\nonumber\\
&=&I(V;Z)-\epsilon_{N},
\end{eqnarray*}
and $\epsilon_{N}\rightarrow 0$ as $N\rightarrow \infty$. From the standard channel coding theorem and the Fano's inequality,
we have (\ref{aox7}).

For the fourth term in (\ref{aox4}), we have
\begin{equation}\label{aox8}
\lim_{N\rightarrow \infty}\frac{1}{N}I(V^{N};Z^{N})\leq I(V;Z),
\end{equation}
and this is from a standard technique as in \cite[p. 343]{CK}.

Substituting (\ref{aox5}), (\ref{aox6}), (\ref{aox7}) and (\ref{aox8}) into (\ref{aox4}), we have
\begin{equation}\label{appe15}
\lim_{N\rightarrow \infty}\Delta\geq I(V;Y)-I(V;Z)=R_{e}.
\end{equation}
$\lim_{N\rightarrow \infty}\Delta\geq R_{e}$ is proved.
Therefore, the achievability proof for Theorem \ref{T5} is completed.

\renewcommand{\theequation}{\arabic{equation}}
\section{Proof of Theorem \ref{T6}\label{appen9}}

Suppose
$(R_{1},R_{2},R_{e})\in \mathcal{R^{G}}$, we will show that $(R_{1},R_{2},R_{e})$ is achievable.
Since $\mathcal{R^{G}}=\mathcal{L}^{(1)}\cup \mathcal{L}^{(2)}\cup \mathcal{L}^{(3)} \cup\mathcal{L}^{(4)}$,
we need to prove that $(R_{1},R_{2},R_{e})\in \mathcal{L}^{(i)}$ ($i=1,2,3,4$) is achievable.

Note that $\mathcal{L}^{(1)}$ is analogous to $\mathcal{L}^{(2)}$, and
$\mathcal{L}^{(3)}$ is analogous to $\mathcal{L}^{(4)}$. Thus, in the remainder of this section,
we only prove that $(R_{1},R_{2},R_{e})\in \mathcal{L}^{(1)}$ and $(R_{1},R_{2},R_{e})\in \mathcal{L}^{(3)}$
are achievable.

\subsection{Achievability of $\mathcal{L}^{(1)}$\label{appen9.1}}

The existence of the encoder-decoder is under the sufficient conditions that
$R_{e}=I(X_{2};Y|X_{1})-I(X_{2};Z|X_{1})+R_{1}$.
Given a triple $(R_{1},R_{2},R_{e})$, choose a joint probability mass function $p_{X_{1},X_{2},Y,Z}(x_{1},x_{2},y,z)$
such that
$$0\leq R_{1}\leq I(X_{1};Y)-I(X_{1};Z), 0\leq R_{2}\leq I(X_{2};Y|X_{1}), R_{1}+R_{2}\leq I(X_{1},X_{2};Y),$$
$$R_{e}\leq R_{1}+R_{2}, R_{e}=I(X_{2};Y|X_{1})-I(X_{2};Z|X_{1})+R_{1}.$$
Note that $R_{e}=I(X_{2};Y|X_{1})-I(X_{2};Z|X_{1})+R_{1}$ implies that
\begin{equation}\label{abc1}
R_{2}\geq I(X_{2};Y|X_{1})-I(X_{2};Z|X_{1}).
\end{equation}
Define
\begin{equation}\label{abc2}
R_{1}=I(X_{1};Y)-I(X_{1};Z)-\gamma,
\end{equation}
where $\gamma\geq 0$.

The confidential message sets $\mathcal{W}_{1}$ and $\mathcal{W}_{2}$ satisfy the following conditions:
\begin{equation}\label{abc3}
\lim_{N\rightarrow \infty}\frac{1}{N}\log\parallel \mathcal{W}_{1}\parallel=R_{1},
\lim_{N\rightarrow \infty}\frac{1}{N}\log\parallel \mathcal{W}_{2}\parallel=R_{2}.
\end{equation}

\textbf{Code-book generation:} Generate $2^{N(I(X_{1};Y)-\gamma-\epsilon_{N})}$ codewords $x_{1}^{N}$ ($\epsilon_{N}\rightarrow \infty$ as $N\rightarrow \infty$),
and each of them is uniformly drawn from the strong typical set
$T^{N}_{X_{1}}(\eta)$. Divide the $2^{N(I(X_{1};Y)-\gamma-\epsilon_{N})}$ codewords into $2^{NR_{1}}$ bins, and each bin corresponds to
a specific value in $\mathcal{W}_{1}$.

Analogously, generate $2^{N(I(X_{2};Y|X_{1})-\epsilon_{N})}$ codewords $x_{2}^{N}$, and each of them is uniformly drawn from the strong typical set
$T^{N}_{X_{2}}(\eta)$. Divide the $2^{N(I(X_{2};Y|X_{1})-\epsilon_{N})}$ codewords into $2^{NR_{2}}$ bins, and each bin corresponds to
a specific value in $\mathcal{W}_{2}$.

\textbf{Decoding scheme:} For a given $y^{N}$, try to find a pair of sequences
$(x_{1}^{N}(\hat{w}_{1}),x_{2}^{N}(\hat{w}_{2}))$ such that
$(x_{1}^{N}(\hat{w}_{1}),x_{2}^{N}(\hat{w}_{2}),y^{N})\in T^{N}_{X_{1}X_{2}Y}(\epsilon)$.
If there exist sequences with the same indices $\hat{w}_{1}$ and $\hat{w}_{2}$, put out the corresponding
$\hat{w}_{1}$ and $\hat{w}_{2}$, else declare a decoding error.

\textbf{Proof of the achievability:}
By using the above definitions, it is easy to verify that  $\lim_{N\rightarrow \infty}\frac{\log\parallel \mathcal{W}_{1}\parallel}{N}= R_{1}$ and
$\lim_{N\rightarrow \infty}\frac{\log\parallel \mathcal{W}_{2}\parallel}{N}= R_{2}$.

Then, note that the codewords of $x_{1}^{N}$ and $x_{2}^{N}$ are respectively upper bounded by $2^{NI(X_{1};Y)}$ and $2^{NI(X_{2};Y|X_{1})}$.
Therefore, from the standard techniques as in \cite[Ch. 14]{cover}, we have $P_{e}\leq \epsilon$.

It remains to show that $\lim_{N\rightarrow \infty}\Delta\geq R_{e}$, see the following.

\begin{eqnarray}\label{abc4}
\lim_{N\rightarrow \infty}\Delta&\triangleq&\lim_{N\rightarrow \infty}\frac{1}{N}H(W_{1},W_{2}|Z^{N})\nonumber\\
&=&\lim_{N\rightarrow \infty}\frac{1}{N}(H(W_{1}|Z^{N})+H(W_{2}|Z^{N},W_{1}))\nonumber\\
&\geq&\lim_{N\rightarrow \infty}\frac{1}{N}(H(W_{1}|Z^{N})+H(W_{2}|Z^{N},W_{1},X_{1}^{N}))\nonumber\\
&\stackrel{(a)}=&\lim_{N\rightarrow \infty}\frac{1}{N}(H(W_{1}|Z^{N})+H(W_{2}|Z^{N},X_{1}^{N}))\nonumber\\
&=&\lim_{N\rightarrow \infty}\frac{1}{N}(H(W_{1},Z^{N})-H(Z^{N})+H(W_{2},Z^{N},X_{1}^{N})-H(Z^{N},X_{1}^{N}))\nonumber\\
&=&\lim_{N\rightarrow \infty}\frac{1}{N}(H(W_{1},Z^{N},X_{1}^{N})-H(X_{1}^{N}|W_{1},Z^{N})-H(Z^{N})\nonumber\\
&&+H(W_{2},Z^{N},X_{1}^{N},X_{2}^{N})-H(X_{2}^{N}|W_{2},Z^{N},X_{1}^{N})-H(Z^{N},X_{1}^{N}))\nonumber\\
&\stackrel{(b)}=&\lim_{N\rightarrow \infty}\frac{1}{N}(H(Z^{N}|X_{1}^{N})+H(W_{1})+H(X_{1}^{N}|W_{1})-H(X_{1}^{N}|W_{1},Z^{N})-H(Z^{N})\nonumber\\
&&+H(Z^{N}|X_{1}^{N},X_{2}^{N})+H(X_{1}^{N})+H(W_{2})+H(X_{2}^{N}|W_{2})-H(X_{2}^{N}|W_{2},Z^{N},X_{1}^{N})-H(Z^{N},X_{1}^{N}))\nonumber\\
&=&\lim_{N\rightarrow \infty}\frac{1}{N}(H(W_{1})+H(X_{1}^{N}|W_{1})-H(X_{1}^{N}|W_{1},Z^{N})+H(X_{2}^{N}|W_{2})+H(W_{2})\nonumber\\
&&-H(X_{2}^{N}|W_{2},Z^{N},X_{1}^{N})-I(X_{1}^{N},X_{2}^{N};Z^{N})),
\end{eqnarray}
where (a) is from $W_{1}\rightarrow (X_{1}^{N},Z^{N})\rightarrow W_{2}$, and (b) is from
$W_{1}\rightarrow X_{1}^{N}\rightarrow Z^{N}$ and $W_{2}\rightarrow (X_{1}^{N},X_{2}^{N})\rightarrow Z^{N}$.

The first term in (\ref{abc4}) is
\begin{equation}\label{abc5}
\lim_{N\rightarrow \infty}\frac{1}{N}H(W_{1})=R_{1}.
\end{equation}

The second term in (\ref{abc4}) can be bounded as follows.
\begin{equation}\label{abc6}
\lim_{N\rightarrow \infty}\frac{1}{N}H(X_{1}^{N}|W_{1})=I(X_{1};Y)-\gamma-R_{1}.
\end{equation}

For the third term in (\ref{abc4}), we have
\begin{equation}\label{abc7}
\lim_{N\rightarrow \infty}\frac{1}{N}H(X_{1}^{N}|W_{1},Z^{N})=0.
\end{equation}
This is because for a given $w_{1}$, there are $2^{N(I(X_{1};Y)-\gamma-\epsilon_{N}-R_{1})}$ codewords left for $x_{1}^{N}$.
Then note that
\begin{eqnarray*}
I(X_{1};Y)-\gamma-\epsilon_{N}-R_{1}&=&I(X_{1};Y)-\gamma-\epsilon_{N}-(I(X_{1};Y)-I(X_{1};Z)-\gamma)\nonumber\\
&=&I(X_{1};Z)-\epsilon_{N},
\end{eqnarray*}
and $\epsilon_{N}\rightarrow 0$ as $N\rightarrow \infty$. From the standard channel coding theorem and the Fano's inequality,
we have (\ref{abc7}).

For the fourth term in (\ref{abc4}), we have
\begin{equation}\label{abc8}
\lim_{N\rightarrow \infty}\frac{1}{N}H(X_{2}^{N}|W_{2})=I(X_{2};Y|X_{1})-R_{2}.
\end{equation}

The fifth term in (\ref{abc4}) is
\begin{equation}\label{abc9}
\lim_{N\rightarrow \infty}\frac{1}{N}H(W_{2})=R_{2}.
\end{equation}

For the sixth term in (\ref{abc4}), we have
\begin{equation}\label{abc10}
\lim_{N\rightarrow \infty}\frac{1}{N}H(X_{2}^{N}|W_{2},Z^{N},X_{1}^{N})=0.
\end{equation}
This is because for given $w_{2}$ and $x_{1}^{N}$, there are $2^{N(I(X_{2};Y|X_{1})-\epsilon_{N}-R_{2})}$ codewords left for $x_{2}^{N}$.
Then note that
\begin{eqnarray*}
I(X_{2};Y|X_{1})-\epsilon_{N}-R_{2}&\stackrel{(1)}\leq&I(X_{2};Y|X_{1})-\epsilon_{N}-(I(X_{2};Y|X_{1})-I(X_{2};Z|X_{1}))\nonumber\\
&=&I(X_{2};Z|X_{1})-\epsilon_{N},
\end{eqnarray*}
where (1) is from (\ref{abc1}), and $\epsilon_{N}\rightarrow 0$ as $N\rightarrow \infty$.
From the standard channel coding theorem and the Fano's inequality,
we have (\ref{abc10}).

For the seventh term in (\ref{abc4}), we have
\begin{equation}\label{abc11}
\lim_{N\rightarrow \infty}\frac{1}{N}I(X_{1}^{N},X_{2}^{N};Z^{N})\leq I(X_{1},X_{2};Z),
\end{equation}
and this is from a standard technique as in \cite[p. 343]{CK}.

Substituting (\ref{abc5}), (\ref{abc6}), (\ref{abc7}), (\ref{abc8}), (\ref{abc9}), (\ref{abc10}) and (\ref{abc11}) into (\ref{abc4}), we have
\begin{eqnarray}\label{abc12}
\lim_{N\rightarrow \infty}\Delta&\geq&I(X_{1};Y)-\gamma+I(X_{2};Y|X_{1})-I(X_{1},X_{2};Z)\nonumber\\
&=&I(X_{1};Y)-\gamma+I(X_{2};Y|X_{1})-I(X_{1};Z)-I(X_{2};Z|X_{1})\nonumber\\
&\stackrel{(a)}=&I(X_{2};Y|X_{1})-I(X_{2};Z|X_{1})+R_{1}=R_{e},
\end{eqnarray}
where (a) is from (\ref{abc2}).

Thus, $\lim_{N\rightarrow \infty}\Delta\geq R_{e}$ is proved.

\subsection{Achievability of $\mathcal{L}^{(3)}$\label{appen9.2}}

The existence of the encoder-decoder is under the sufficient conditions that
$R_{e}=I(X_{1},X_{2};Y)-I(X_{1},X_{2};Z)$.
Given a triple $(R_{1},R_{2},R_{e})$, choose a joint probability mass function $p_{X_{1},X_{2},Y,Z}(x_{1},x_{2},y,z)$
such that
$$I(X_{1};Y)-I(X_{1};Z)\leq R_{1}\leq I(X_{1};Y), I(X_{2};Y|X_{1})-I(X_{2};Z|X_{1})\leq R_{2}\leq I(X_{2};Y|X_{1}),
R_{1}+R_{2}\leq I(X_{1},X_{2};Y),$$
$$R_{e}\leq R_{1}+R_{2}, \ \ R_{e}=I(X_{1},X_{2};Y)-I(X_{1},X_{2};Z).$$

The confidential message sets $\mathcal{W}_{1}$ and $\mathcal{W}_{2}$ satisfy the following conditions:
\begin{equation}\label{abcx3}
\lim_{N\rightarrow \infty}\frac{1}{N}\log\parallel \mathcal{W}_{1}\parallel=R_{1},
\lim_{N\rightarrow \infty}\frac{1}{N}\log\parallel \mathcal{W}_{2}\parallel=R_{2}.
\end{equation}

\textbf{Code-book generation:} Generate $2^{N(I(X_{1};Y)-\epsilon_{N})}$ codewords $x_{1}^{N}$ ($\epsilon_{N}\rightarrow \infty$ as $N\rightarrow \infty$),
and each of them is uniformly drawn from the strong typical set
$T^{N}_{X_{1}}(\eta)$. Divide the $2^{N(I(X_{1};Y)-\epsilon_{N})}$ codewords into $2^{NR_{1}}$ bins, and each bin corresponds to
a specific value in $\mathcal{W}_{1}$.

Analogously, generate $2^{N(I(X_{2};Y|X_{1})-\epsilon_{N})}$ codewords $x_{2}^{N}$, and each of them is uniformly drawn from the strong typical set
$T^{N}_{X_{2}}(\eta)$. Divide the $2^{N(I(X_{2};Y|X_{1})-\epsilon_{N})}$ codewords into $2^{NR_{2}}$ bins, and each bin corresponds to
a specific value in $\mathcal{W}_{2}$.

\textbf{Decoding scheme:} For a given $y^{N}$, try to find a pair of sequences
$(x_{1}^{N}(\hat{w}_{1}),x_{2}^{N}(\hat{w}_{2}))$ such that
$(x_{1}^{N}(\hat{w}_{1}),x_{2}^{N}(\hat{w}_{2}),y^{N})\in T^{N}_{X_{1}X_{2}Y}(\epsilon)$.
If there exist sequences with the same indices $\hat{w}_{1}$ and $\hat{w}_{2}$, put out the corresponding
$\hat{w}_{1}$ and $\hat{w}_{2}$, else declare a decoding error.

\textbf{Proof of the achievability:}
By using the above definitions, it is easy to verify that  $\lim_{N\rightarrow \infty}\frac{\log\parallel \mathcal{W}_{1}\parallel}{N}= R_{1}$ and
$\lim_{N\rightarrow \infty}\frac{\log\parallel \mathcal{W}_{2}\parallel}{N}= R_{2}$.

Then, note that the codewords of $x_{1}^{N}$ and $x_{2}^{N}$ are respectively upper bounded by $2^{NI(X_{1};Y|X_{2})}$ and $2^{NI(X_{2};Y|X_{1})}$.
Therefore, from the standard techniques as in \cite[Ch. 14]{cover}, we have $P_{e}\leq \epsilon$.

It remains to show that $\lim_{N\rightarrow \infty}\Delta\geq R_{e}$, see the following.

\begin{eqnarray}\label{abcx4}
\lim_{N\rightarrow \infty}\Delta&\triangleq&\lim_{N\rightarrow \infty}\frac{1}{N}H(W_{1},W_{2}|Z^{N})\nonumber\\
&=&\lim_{N\rightarrow \infty}\frac{1}{N}(H(W_{1}|Z^{N})+H(W_{2}|Z^{N},W_{1}))\nonumber\\
&\geq&\lim_{N\rightarrow \infty}\frac{1}{N}(H(W_{1}|Z^{N})+H(W_{2}|Z^{N},W_{1},X_{1}^{N}))\nonumber\\
&\stackrel{(a)}=&\lim_{N\rightarrow \infty}\frac{1}{N}(H(W_{1}|Z^{N})+H(W_{2}|Z^{N},X_{1}^{N}))\nonumber\\
&=&\lim_{N\rightarrow \infty}\frac{1}{N}(H(W_{1},Z^{N})-H(Z^{N})+H(W_{2},Z^{N},X_{1}^{N})-H(Z^{N},X_{1}^{N}))\nonumber\\
&=&\lim_{N\rightarrow \infty}\frac{1}{N}(H(W_{1},Z^{N},X_{1}^{N})-H(X_{1}^{N}|W_{1},Z^{N})-H(Z^{N})\nonumber\\
&&+H(W_{2},Z^{N},X_{1}^{N},X_{2}^{N})-H(X_{2}^{N}|W_{2},Z^{N},X_{1}^{N})-H(Z^{N},X_{1}^{N}))\nonumber\\
&\stackrel{(b)}=&\lim_{N\rightarrow \infty}\frac{1}{N}(H(Z^{N}|X_{1}^{N})+H(W_{1})+H(X_{1}^{N}|W_{1})-H(X_{1}^{N}|W_{1},Z^{N})-H(Z^{N})\nonumber\\
&&+H(Z^{N}|X_{1}^{N},X_{2}^{N})+H(X_{1}^{N})+H(W_{2})+H(X_{2}^{N}|W_{2})-H(X_{2}^{N}|W_{2},Z^{N},X_{1}^{N})-H(Z^{N},X_{1}^{N}))\nonumber\\
&=&\lim_{N\rightarrow \infty}\frac{1}{N}(H(W_{1})+H(X_{1}^{N}|W_{1})-H(X_{1}^{N}|W_{1},Z^{N})+H(X_{2}^{N}|W_{2})+H(W_{2})\nonumber\\
&&-H(X_{2}^{N}|W_{2},Z^{N},X_{1}^{N})-I(X_{1}^{N},X_{2}^{N};Z^{N})),
\end{eqnarray}
where (a) is from $W_{1}\rightarrow (X_{1}^{N},Z^{N})\rightarrow W_{2}$, and (b) is from
$W_{1}\rightarrow X_{1}^{N}\rightarrow Z^{N}$ and $W_{2}\rightarrow (X_{1}^{N},X_{2}^{N})\rightarrow Z^{N}$.

The first term in (\ref{abcx4}) is
\begin{equation}\label{abcx5}
\lim_{N\rightarrow \infty}\frac{1}{N}H(W_{1})=R_{1}.
\end{equation}

The second term in (\ref{abcx4}) can be bounded as follows.
\begin{equation}\label{abcx6}
\lim_{N\rightarrow \infty}\frac{1}{N}H(X_{1}^{N}|W_{1})=I(X_{1};Y)-R_{1}.
\end{equation}

For the third term in (\ref{abcx4}), we have
\begin{equation}\label{abcx7}
\lim_{N\rightarrow \infty}\frac{1}{N}H(X_{1}^{N}|W_{1},Z^{N})=0.
\end{equation}
This is because for a given $w_{1}$, there are $2^{N(I(X_{1};Y)-\epsilon_{N}-R_{1})}$ codewords left for $x_{1}^{N}$.
Then note that
\begin{eqnarray*}
I(X_{1};Y)-\epsilon_{N}-R_{1}&\leq&I(X_{1};Y)-\epsilon_{N}-(I(X_{1};Y)-I(X_{1};Z))\nonumber\\
&=&I(X_{1};Z)-\epsilon_{N},
\end{eqnarray*}
and $\epsilon_{N}\rightarrow 0$ as $N\rightarrow \infty$. From the standard channel coding theorem and the Fano's inequality,
we have (\ref{abcx7}).

For the fourth term in (\ref{abcx4}), we have
\begin{equation}\label{abcx8}
\lim_{N\rightarrow \infty}\frac{1}{N}H(X_{2}^{N}|W_{2})=I(X_{2};Y|X_{1})-R_{2}.
\end{equation}

The fifth term in (\ref{abcx4}) is
\begin{equation}\label{abcx9}
\lim_{N\rightarrow \infty}\frac{1}{N}H(W_{2})=R_{2}.
\end{equation}

For the sixth term in (\ref{abcx4}), we have
\begin{equation}\label{abcx10}
\lim_{N\rightarrow \infty}\frac{1}{N}H(X_{2}^{N}|W_{2},Z^{N},X_{1}^{N})=0.
\end{equation}
This is because for given $w_{2}$ and $x_{1}^{N}$, there are $2^{N(I(X_{2};Y|X_{1})-\epsilon_{N}-R_{2})}$ codewords left for $x_{2}^{N}$.
Then note that
\begin{eqnarray*}
I(X_{2};Y|X_{1})-\epsilon_{N}-R_{2}&\leq&I(X_{2};Y|X_{1})-\epsilon_{N}-(I(X_{2};Y|X_{1})-I(X_{2};Z|X_{1}))\nonumber\\
&=&I(X_{2};Z|X_{1})-\epsilon_{N},
\end{eqnarray*}
and $\epsilon_{N}\rightarrow 0$ as $N\rightarrow \infty$.
From the standard channel coding theorem and the Fano's inequality,
we have (\ref{abcx10}).

For the seventh term in (\ref{abcx4}), we have
\begin{equation}\label{abcx11}
\lim_{N\rightarrow \infty}\frac{1}{N}I(X_{1}^{N},X_{2}^{N};Z^{N})\leq I(X_{1},X_{2};Z),
\end{equation}
and this is from a standard technique as in \cite[p. 343]{CK}.

Substituting (\ref{abcx5}), (\ref{abcx6}), (\ref{abcx7}), (\ref{abcx8}), (\ref{abcx9}), (\ref{abcx10}) and (\ref{abcx11}) into (\ref{abcx4}), we have
\begin{eqnarray}\label{abcx12}
\lim_{N\rightarrow \infty}\Delta&\geq&I(X_{1};Y)+I(X_{2};Y|X_{1})-I(X_{1},X_{2};Z)\nonumber\\
&=&I(X_{1},X_{2};Y)-I(X_{1},X_{2};Z)=R_{e}.
\end{eqnarray}

Thus, $\lim_{N\rightarrow \infty}\Delta\geq R_{e}$ is proved.

\renewcommand{\theequation}{\arabic{equation}}
\section{Proof of Theorem \ref{T7}\label{appen10}}

In this section, we prove Theorem \ref{T7}. The first three bounds in Theorem \ref{T7} are the capacity region the MAC,
and the proof is omitted. It remains to prove $R_{e}\leq R_{1}+R_{2}$ and $R_{e}\leq I(X_{1},X_{2};Y)-I(X_{1},X_{2};Z)$, see
the followings.

\textbf{(Proof of $R_{e}\leq R_{1}+R_{2}$)} The inequality is obtained by the following equation.
\begin{equation}\label{axa1}
R_{e}\leq\lim_{N\rightarrow \infty}\frac{1}{N}H(W_{1},W_{2}|Z^{N})\leq \lim_{N\rightarrow \infty}\frac{1}{N}H(W_{1},W_{2})=R_{1}+R_{2}.
\end{equation}

\textbf{(Proof of $R_{e}\leq I(X_{1},X_{2};Y)-I(X_{1},X_{2};Z)$)} The proof is obtained by the following (\ref{axa5}).
\begin{eqnarray}\label{axa5}
R_{e}&\leq&\lim_{N\rightarrow \infty}\frac{1}{N}H(W_{1},W_{2}|Z^{N})\nonumber\\
&\stackrel{(a)}\leq&\lim_{N\rightarrow \infty}\frac{1}{N}(H(W_{1},W_{2}|Z^{N})+\delta(P_{e})-H(W_{1},W_{2}|Z^{N},Y^{N}))\nonumber\\
&=&\lim_{N\rightarrow \infty}\frac{1}{N}(I(W_{1},W_{2};Y^{N}|Z^{N})+\delta(P_{e}))\nonumber\\
&=&\lim_{N\rightarrow \infty}\frac{1}{N}(H(Y^{N}|Z^{N})-H(Y^{N}|Z^{N},W_{1},W_{2})+\delta(P_{e}))\nonumber\\
&\leq&\lim_{N\rightarrow \infty}\frac{1}{N}(H(Y^{N}|Z^{N})-H(Y^{N}|Z^{N},W_{1},W_{2},X_{1}^{N},X_{2}^{N})+\delta(P_{e}))\nonumber\\
&\stackrel{(b)}=&\lim_{N\rightarrow \infty}\frac{1}{N}(H(Y^{N}|Z^{N})-H(Y^{N}|Z^{N},X_{1}^{N},X_{2}^{N})+\delta(P_{e}))\nonumber\\
&=&\lim_{N\rightarrow \infty}\frac{1}{N}(I(Y^{N};X_{1}^{N},X_{2}^{N}|Z^{N})+\delta(P_{e}))\nonumber\\
&=&\lim_{N\rightarrow \infty}\frac{1}{N}(I(Y^{N};X_{1}^{N},X_{2}^{N})-I(Z^{N};X_{1}^{N},X_{2}^{N})+\delta(P_{e}))\nonumber\\
&=&\lim_{N\rightarrow \infty}(\frac{1}{N}\sum_{i=1}^{N}(H(Y_{i}|Y^{i-1})-H(Y_{i}|X_{1,i},X_{2,i})\nonumber\\
&&-H(Z_{i}|Z^{i-1})+H(Z_{i}|X_{1,i},X_{2,i}))+\frac{1}{N}\delta(P_{e}))\nonumber\\
&\leq&\lim_{N\rightarrow \infty}(\frac{1}{N}\sum_{i=1}^{N}(H(Y_{i}|Y^{i-1})-H(Y_{i}|X_{1,i},X_{2,i})\nonumber\\
&&-H(Z_{i}|Z^{i-1},Y^{i-1})+H(Z_{i}|X_{1,i},X_{2,i}))+\frac{1}{N}\delta(P_{e}))\nonumber\\
&\stackrel{(c)}=&\lim_{N\rightarrow \infty}(\frac{1}{N}\sum_{i=1}^{N}(H(Y_{i}|Y^{i-1})-H(Y_{i}|X_{1,i},X_{2,i})\nonumber\\
&&-H(Z_{i}|Y^{i-1})+H(Z_{i}|X_{1,i},X_{2,i}))+\frac{1}{N}\delta(P_{e}))\nonumber\\
&\stackrel{(d)}\leq&\lim_{N\rightarrow \infty}(\frac{1}{N}\sum_{i=1}^{N}(H(Y_{i})-H(Y_{i}|X_{1,i},X_{2,i})\nonumber\\
&&-H(Z_{i})+H(Z_{i}|X_{1,i},X_{2,i}))+\frac{1}{N}\delta(P_{e}))\nonumber\\
&\stackrel{(e)}\leq&\lim_{N\rightarrow \infty}(H(Y)-H(Y|X_{1},X_{2})\nonumber\\
&&-H(Z)+H(Z|X_{1},X_{2})+\frac{1}{N}\delta(P_{e}))\nonumber\\
&=&\lim_{N\rightarrow \infty}(I(X_{1},X_{2};Y)-I(X_{1},X_{2};Z)+\frac{1}{N}\delta(P_{e}))\nonumber\\
&\stackrel{(f)}\leq&\lim_{N\rightarrow \infty}(I(X_{1},X_{2};Y)-I(X_{1},X_{2};Z)+\frac{1}{N}\delta(\epsilon))\nonumber\\
&=&I(X_{1},X_{2};Y)-I(X_{1},X_{2};Z),
\end{eqnarray}
where (a) is from the Fano's inequality, and (b) is from $(W_{1},W_{2})\rightarrow (Z^{N},X_{2}^{N},X_{1}^{N})\rightarrow Y^{N}$,
(c) is from $Z^{i-1}\rightarrow Y^{i-1}\rightarrow Z_{i}$, (d) is from $Y^{i-1}\rightarrow Y_{i}\rightarrow Z_{i}$,
(e) is from the definitions that $X_{1}\triangleq (X_{1,J}, J)$, $X_{2}\triangleq (X_{2,J}, J)$, $Y\triangleq Y_{J}$, $Z\triangleq Z_{J}$
where $J$ is a random variable (uniformly distributed over $\{1,2,...,N\}$), and it is independent of $X_{1,i}$, $X_{2,i}$, $Y_{i}$ and $Z_{i}$,
and (f) is from $P_{e}\leq \epsilon$.

The proof of Theorem \ref{T7} is completed.

\renewcommand{\theequation}{\arabic{equation}}
\section{Proof of Theorem \ref{T9}\label{appen11}}

Theorem \ref{T9} is proved by calculating the mutual information terms in $\mathcal{C}_{s}^{(G)}$ and $\mathcal{C}_{s}^{(H)}$,
see the following.

All the random variables take values in $\{0, 1\}$.
Let $Pr\{X_{1}=0\}=\alpha$, $Pr\{X_{1}=1\}=1-\alpha$, $Pr\{X_{2}=0\}=\beta$ and $Pr\{X_{2}=1\}=1-\beta$.
Note that $X_{1}$, $X_{2}$, $Y$ and $Z$ satisfy
\begin{equation}\label{appeeex1}
Y=X_{1}\cdot X_{2}, \ \ Z=Y\oplus Z^{*},
\end{equation}
where $X_{1}$ is independent of $X_{2}$, and $Pr\{Z^{*}=0\}=1-p$, $Pr\{Z^{*}=1\}=p$.

The joint probability $p_{X_{1}X_{2}Y}$ is calculated by the following (\ref{appeeex2}).
\begin{equation}\label{appeeex2}
p_{X_{1},X_{2},Y}(x_{1},x_{2},y)=p_{Y|X_{1},X_{2}}(y|x_{1},x_{2})p_{X_{1}}(x_{1})p_{X_{2}}(x_{2}).
\end{equation}
The joint probability $p_{X_{1}X_{2}Z}$ is calculated by the following (\ref{appeeex3}).
\begin{equation}\label{appeeex3}
p_{X_{1},X_{2},Z}(x_{1},x_{2},z)=\sum_{y}p_{Z|Y}(z|y)p_{Y|X_{1},X_{2}}(y|x_{1},x_{2})p_{X_{1}}(x_{1})p_{X_{2}}(x_{2}).
\end{equation}

Then, $\mathcal{C}_{s}^{(G)}$ is
\begin{equation}
\mathcal{C}_{s}^{(G)}=\left\{
\begin{array}{ll}
(R_{1}, R_{2}):
\begin{array}{ll}
0\leq R_{1}\leq h(p)\\
0\leq R_{2}\leq h(p)\\
R_{1}+R_{2}\leq h(p)
\end{array}
\end{array}
\right\}.
\end{equation}

Moreover, $\mathcal{C}_{s}^{(H)}$ is

\begin{equation}
\mathcal{C}_{s}^{(H)}=\left\{
\begin{array}{ll}
(R_{1}, R_{2}):
\begin{array}{ll}
0\leq R_{1}\leq 1\\
0\leq R_{2}\leq 1\\
R_{1}+R_{2}\leq h(p)
\end{array}
\end{array}
\right\}.
\end{equation}

It is easy to see that $\mathcal{C}_{s}^{(G)}$ and $\mathcal{C}_{s}^{(H)}$ are the same for the binary case, and therefore,
the secrecy capacity region for the binary case of degraded MAC-WT with non-cooperative encoders is
\begin{equation}
\mathcal{R}^{(J)}=\left\{
\begin{array}{ll}
(R_{1}, R_{2}):
\begin{array}{ll}
0\leq R_{1}\leq h(p)\\
0\leq R_{2}\leq h(p)\\
R_{1}+R_{2}\leq h(p)
\end{array}
\end{array}
\right\}.
\end{equation}


\begin{thebibliography}{99}

\bibitem{Wy} A. D. Wyner, ``The wire-tap channel,"
{\sl The Bell System Technical Journal}, vol. 54, no. 8, pp.
1355-1387, 1975.


\bibitem{CK} I. Csisz$\acute{a}$r and J. K\"{o}rner, ``Broadcast channels with confidential messages," {\sl IEEE Trans
Inf Theory}, vol. IT-24, no. 3, pp. 339-348, May 1978.

\bibitem{KM} J. K\"{o}rner and K. Marton, ``General broadcast channels with degraded message sets,"
{\sl IEEE Trans Inf Theory}, vol. IT-23, no. 1, pp. 60-64, January 1977.

\bibitem{CH} S. K. Leung-Yan-Cheong, M. E. Hellman,
``The Gaussian wire-tap channel," {\sl IEEE Trans Inf Theory}, vol. IT-24, no. 4, pp. 451-456, July 1978.

\bibitem{AC} R. Ahlswede and N. Cai, ``Transmission, Identification and Common Randomness Capacities for
Wire-Tap Channels with Secure Feedback from the Decoder," book
chapter in {\sl General Theory of Information Transfer and
Combinatorics}, LNCS 4123,  pp. 258-275, Berlin: Springer-Verlag,
2006.

\bibitem{AFJK} E. Ardestanizadeh, M. Franceschetti, T.Javidi and Y.Kim, ``Wiretap channel with secure rate-limited feedback,"
{\sl IEEE Trans Inf Theory}, vol. IT-55, no. 12, pp. 5353-5361,
December 2009.

\bibitem{LGP} L. Lai, H. El Gamal and V. Poor,
``The wiretap channel with feedback: encryption over the channel,"
{\sl IEEE Trans Inf Theory}, vol. IT-54, pp. 5059-5067, 2008.

\bibitem{Me} N. Merhav, ``Shannon's secrecy system with informed receivers and its application to systematic coding
for wiretapped channels," {\sl IEEE Trans Inf Theory}, special issue
on Information-Theoretic Security, vol. IT-54, no. 6, pp. 2723-2734,
June 2008.


\bibitem{Ch} Y. Chen, A. J. Han Vinck, ``Wiretap channel with side information,"
{\sl IEEE Trans Inf Theory}, vol. IT-54, no. 1, pp. 395-402, January 2008.

\bibitem{MVL} C. Mitrpant, A. J. Han Vinck and Y. Luo,
``An Achievable Region for the Gaussian Wiretap Channel with Side
Information," {\sl IEEE Trans Inf Theory}, vol. IT-52, no. 5, pp.
2181-2190, 2006.

\bibitem{LG2} L. Lai and H. El Gamal, ¡°The relay-eavesdropper channel: cooperation
for secrecy,¡± {\sl IEEE Trans Inf Theory}, vol. IT-54, no. 9, pp. 4005¨C4019,
Sep. 2008.

\bibitem{LMSY} R. Liu, I. Maric, P. Spasojevic and R.D Yates, ¡°Discrete memoryless
interference and broadcast channels with confidential messages: secrecy
rate regions,¡± {\sl IEEE Trans Inf Theory}, vol. IT-54, no. 6, pp. 2493-2507, Jun.
2008.

\bibitem{LP} Y. Liang and H. V. Poor, ¡°Multiple-access channels with confidential
messages," {\sl IEEE Trans Inf Theory}, vol. IT-54, no. 3, pp. 976-1002,
Mar. 2008.

\bibitem{TY2} E. Tekin and A. Yener, ¡°The Gaussian multiple access wire-tap channel,¡±
{\sl IEEE Trans Inf Theory}, vol. IT-54, no. 12, pp. 5747-5755, Dec. 2008.


\bibitem{TY1} E. Tekin and A. Yener, ¡°The general Gaussian multiple access and
two-way wire-tap channels: Achievable rates and cooperative jamming,¡±
{\sl IEEE Trans Inf Theory}, vol. IT-54, no. 6, pp. 2735-2751, June 2008.


\bibitem{EU} E. Ekrem and S. Ulukus, ¡°On the secrecy of multiple access wiretap
channel,¡± in {\sl Proc. Annual Allerton Conf. on Communications, Control
and Computing}, Monticello, IL, Sept. 2008.

\bibitem{Cs} I. Csisz$\acute{a}$r and J. K\"{o}rner, {\sl Information Theory. Coding Theorems for Discrete Memoryless Systems.}
London, U.K.: Academic, 1981.

\bibitem{cover} T. M. Cover and J. A. Thomas, {\sl Elements of Information Theory.}
New York: Wiley, 1991.




\end{thebibliography}
\end{document}